\documentclass[acmsmall, screen, nonacm]{acmart}

\title{Resource Estimation for Fault-Tolerant Quantum Programs}

\author{Bonan Su} 
\orcid{0009-0009-7279-0658}
\email{sbn24@mails.tsinghua.edu.cn}
\affiliation{
    \department{Department of Computer Science and Technology}
    \institution{Tsinghua University}
    \city{Beijing}
    \country{China}
}

\author{Yuan Feng}
\orcid{0000-0002-3097-3896}
\affiliation{
    \institution{Tsinghua University}
    \city{Beijing}
    \country{China}
}
\email{yuan_feng@tsinghua.edu.cn}

\author{Li Zhou}
\orcid{0000-0002-9868-8477}
\affiliation{
    \institution{Institute of Software, Chinese Academy of Sciences}
    \city{Beijing}
    \country{China}
}
\email{zhouli@ios.ac.cn}

\author{Mingsheng Ying}
\orcid{0000-0003-4847-702X}
\affiliation{%
  \institution{University of Technology Sydney}
  \city{Sydney}
  \country{Australia}
}
\email{mingsheng.ying@uts.edu.au}

\usepackage{mathrsfs}
\usepackage{mathpartir}
\usepackage{extarrows}
\usepackage{cancel}
\usepackage{makecell}
\usepackage{stmaryrd}
\usepackage{lineno}
\usepackage{subcaption}
\usepackage{tcolorbox}
\usepackage{enumitem}
\usepackage{booktabs}
\usepackage{longtable}
\usepackage{hyperref}
\usepackage{pgfplots}
\usepgfplotslibrary{groupplots}
\pgfplotsset{compat=1.18}

\usepackage{natbib}

\usepackage{amssymb}
\usepackage{listings}
\usepackage{tikz}
\usetikzlibrary{quantikz2}
\usetikzlibrary{arrows.meta}

\newcommand{\bR}{\mathbb{R}}
\newcommand{\bN}{\mathbb{N}}

\newcommand{\cS}{\mathcal{S}}

\newcommand{\cB}{\mathcal{B}}
\newcommand{\cC}{\mathcal{C}}

\newcommand{\cM}{\mathcal{M}}

\newcommand{\dom}{\mathrm{dom}}

\newcommand{\cnorm}[1]{{\left\vert\kern-0.25ex\left\vert\kern-0.25ex\left\vert #1 \right\vert\kern-0.25ex\right\vert\kern-0.25ex\right\vert}}
\renewcommand{\>}{\rangle}
\newcommand{\<}{\langle}
\renewcommand{\bar}[1]{\overline{#1}}
\newcommand{\pare}[1]{\left ( #1 \right )}

\definecolor{darkgreen}{rgb}{0.0, 0.4, 0.0}

\newcommand{\sem}[1]{\left \llbracket #1 \right \rrbracket}

\newcommand{\tif}{\mathbf{if}}
\newcommand{\tthen}{\mathbf{then}}

\newcommand{\tskip}{\mathbf{skip}}

\newcommand{\trelease}{\mathbf{release}}
\newcommand{\tlogical}{\mathbf{logical}}
\newcommand{\treset}{\mathbf{prepare}}
\newcommand{\tnew}{\mathbf{new}}
\newcommand{\tcall}{\mathbf{call}}

\newcommand{\tdef}{\mathbf{def}}

\newcommand{\ActBlock}{\mathrm{ActBlock}}

\newcommand{\rCNOT}{\mathrm{CNOT}}

\newcommand{\yellowbox}{\tikz[baseline=-0.5ex]{\node[draw, rounded corners=3pt, fill=yellow!20, minimum width=1em, minimum height=2ex] {};}}

\usepackage{xcolor}
\definecolor{lightblue}{RGB}{233,242,246}
\definecolor{darkblue}{RGB}{230,230,253}
\definecolor{ddarkblue}{RGB}{11,0,114}

\newcommand{\bluebox}{\tikz[baseline=-0.5ex]{\node[draw, fill=lightblue, minimum width=1ex, minimum height=2ex] {};}}
\newcommand{\dbluebox}{\tikz[baseline=-0.5ex]{\node[draw=ddarkblue, line width=0.8pt, fill=darkblue, minimum width=1em, minimum height=2ex] {};}}

\newcommand{\gradbox}{
\tikz[baseline=-0.5ex]{
    \node[
        draw,
        line width=0pt,
        left color=red!23,
        right color=green!10,
        minimum width=2em,
        minimum height=1.5ex
    ] {};
}
}

\usepackage{cleveref}
\usepackage{amsthm}
\usepackage{wrapfig}
\usepackage{tabularx}
\usepackage{booktabs}

\theoremstyle{plain} 

\newtheorem{theorem}{Theorem}[section]

\newtheorem{example}[theorem]{Example}

\definecolor{ftqpKeyword}{HTML}{1D4ED8}
\definecolor{ftqpType}{HTML}{7C3AED}
\definecolor{ftqpField}{HTML}{047857}
\definecolor{ftqpPauli}{HTML}{B45309}
\definecolor{ftqpComment}{HTML}{64748B}
\definecolor{ftqpString}{HTML}{BE123C}
\definecolor{ftqpNumber}{HTML}{0F766E}
\definecolor{ftqpBg}{HTML}{F8FAFC}

\lstdefinelanguage{FTQP}{
sensitive=true,
alsoletter={\#},
morekeywords={
    hardware,qecc,using,on,def,program,
    repeat,until,
    logical,new,release,prepare,reset,if,then,call,skip,parallel
},
morekeywords=[2]{Pauli,Codeblock,CodeBlock,Block},
morekeywords=[3]{SE,PREP,OPL,OPF,arity,time,tSE,tdec,pL,pF,naux,n_aux,\#SE,
\#ancilla},
morekeywords=[4]{I,X,Y,Z,M,H,ZZ},
morecomment=[l]{//},
morecomment=[l]{\#},
morecomment=[s]{/*}{*/},
morestring=[b]",
literate=
    {:=}{{{\color{ftqpKeyword}:=}}}2
    {||}{{{\color{ftqpKeyword}||}}}2
    {->}{{{\color{ftqpKeyword}->}}}2
    {<=}{{{\color{ftqpKeyword}<=}}}2
    {>=}{{{\color{ftqpKeyword}>=}}}2
    {==}{{{\color{ftqpKeyword}==}}}2
    {!=}{{{\color{ftqpKeyword}!=}}}2
}

\lstdefinestyle{ftqp}{
language=FTQP,
basicstyle=\ttfamily\tiny,
keywordstyle=\color{ftqpKeyword}\bfseries,
keywordstyle=[2]\color{ftqpType}\bfseries,
keywordstyle=[3]\color{ftqpField},
keywordstyle=[4]\color{ftqpPauli}\bfseries,
commentstyle=\color{ftqpComment}\itshape,
stringstyle=\color{ftqpString},
numberstyle=\tiny\color{ftqpComment},
xleftmargin=1.5em,
xrightmargin=0.5em,
numbers=left,
stepnumber=1,
numbersep=8pt,
backgroundcolor=\color{ftqpBg},
frame=single,
rulecolor=\color{black!15},
breaklines=true,
columns=fullflexible,
keepspaces=true,
showstringspaces=false,
tabsize=2
}

\lstdefinelanguage{bash}{
  keywords={cd, ls, echo, cat, grep, awk, sed, sudo, chmod, chown, exit, export, unset, source, alias, history, pwd, kill, ps, top, make},
  keywordstyle=\color{blue}\bfseries,
  ndkeywords={if, then, else, fi, for, in, do, done, while, case, esac, function},
  ndkeywordstyle=\color{purple}\bfseries,
  identifierstyle=\color{black},
  sensitive=false,
  comment=[l]{\#},
  commentstyle=\color{gray}\ttfamily,
  stringstyle=\color{teal},
  morestring=[b]',
  morestring=[b]"
}

\lstdefinestyle{bashstyle}{
  language=bash,
  basicstyle=\ttfamily\small,
  backgroundcolor=\color{gray!5},
  frame=single,
  rulecolor=\color{gray!40},
  breaklines=true,
  breakatwhitespace=true,
  showstringspaces=false,
  columns=fullflexible,
  keepspaces=true,
  tabsize=2,
  captionpos=b,
  xleftmargin=1em,
  xrightmargin=1em
}

\lstdefinestyle{output}{
basicstyle=\ttfamily\small,
xleftmargin=1.5em,
xrightmargin=0.5em,
backgroundcolor=\color{ftqpBg},
frame=single,
rulecolor=\color{black!15},
breaklines=true,
columns=fullflexible,
keepspaces=true,
showstringspaces=false,
tabsize=2
}

\allowdisplaybreaks

\begin{document}

\begin{abstract}
Fault-tolerant quantum computation enables the deployment of practical quantum algorithms but incurs substantial overhead from error correction, making resource estimation a central concern. 
Beyond case-by-case analyses, existing quantum programming languages either require programmers to manipulate low-level hardware details, rendering fault-tolerant implementations cumbersome, or abstract away the underlying error-correction schemes, reducing the effectiveness of resource utilization and estimation.

To address these limitations while preserving programmability, we present a quantum programming language that enables efficient resource utilization, together with a resource-estimation framework for comprehensive resource analysis.
Our framework features programmer-visible abstractions of error-correction schemes and cross-layer program-hardware analysis, allowing systematic exploration of resource trade-offs.
We evaluate our approach on detailed fault-tolerant implementations of practical large-scale quantum algorithms, including components typically treated as black boxes in existing frameworks. 
The results demonstrate that our framework enables substantial resource savings while delivering detailed, fine-grained, and accurate resource estimates for fault-tolerant quantum programs.
\end{abstract}

\maketitle

\section{Introduction}\label{sec:intro}


To realize practical quantum algorithms capable of solving real-world problems, there is a growing consensus that the field should move beyond the noisy intermediate-scale quantum (NISQ) era and toward fault-tolerant quantum computation (FTQC)~\cite{Preskill2018quantumcomputingin,Campbell2017roadstowards}, where logical qubits are encoded using multiple physical qubits via quantum \emph{error-correction schemes}, and fault-tolerant operations are implemented through interleavings of physical operations, syndrome extraction, classical decoding, and corrective procedures.

This incurs substantial resource overhead and introduces a gap between fault-tolerant implementations and the underlying physical resource consumption.
Bridging this gap to estimate the physical resources required by fault-tolerant implementations has become
a compelling topic and naturally raises challenges for programming languages: rather than relying on case-by-case analyses~\cite{babbush2026securingellipticcurvecryptocurrencies,cain2026shorsalgorithmpossible10000}, how can programming support enable more efficient resource utilization and more comprehensive resource estimation?

To answer this question, existing quantum programming languages tend to follow two contrasting approaches.
On the one hand, languages targeting practical NISQ devices, such as Qiskit~\cite{qiskit}, treat program variables as physical qubits and naturally expose resource consumption through compilation and transpilation.
However, the absence of suitable abstractions for error-correction schemes makes fault-tolerant programming highly cumbersome.
On the other hand, high-level languages that provide a logical programming abstraction, such as Microsoft's Q\#~\cite{microsoft_resource_estimator,assessing}, abstract away the underlying error-correction schemes from programmers.
As a result, although programmability is improved, resource efficiency and the faithfulness of resource estimation to practical implementations may be compromised.

These two approaches are analogous to assembly language and high-level languages such as Python in classical programming: the former operate close to the hardware, whereas the latter abstract away low-level details.
To strike a balance between them, we present a quantum programming language together with a corresponding resource-estimation framework that selectively exposes fault-tolerant interfaces across abstraction layers, enabling more efficient resource utilization and more realistic resource estimation.
This contrasts with existing resource estimators, where error-correction schemes are typically modeled through backend parameters rather than exposed as programmer-visible abstractions~\cite{microsoft_resource_estimator,zapata_benchq,obenland_2026_18154991,harrigan2024expressinganalyzingquantumalgorithms,rigetti,bärtschi2025potentialapplicationsquantumcomputing,Leblond_2024}.

In the following, we present the key design features of our framework, followed by a discussion of their benefits, the challenges they introduce, and our corresponding solutions for each component in turn, based on the modular architecture of our resource-estimation framework shown in Fig.~\ref{fig:architecture}.
\begin{figure}[t]
    \centering
    \includegraphics[width=\textwidth]{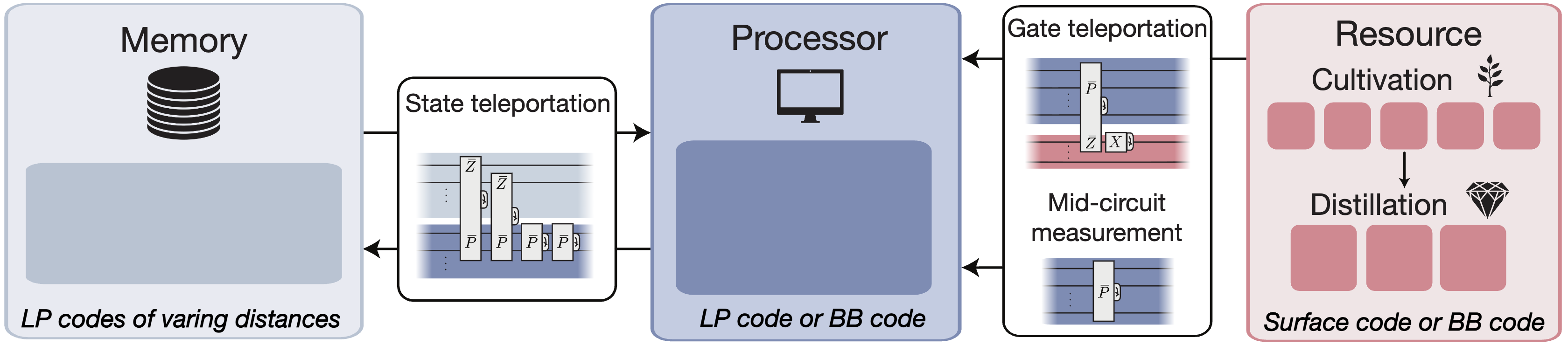}
    \caption{Error-correction schemes assigned to different functional modules in an implementation of Shor's algorithm, where LP, BB, and surface codes represent error-correction schemes.
    Adapted from Fig.~2b of~\cite{cain2026shorsalgorithmpossible10000}.}
    \label{fig:shor}
\end{figure}
\paragraph{Core Design Principle: Associating Logical Qubits with Error-Correction Schemes}
To facilitate fault-tolerant quantum computation, our language operates at the logical level and uses program variables to address logical qubits.
However, rather than completely abstracting away the underlying information, our language associates each logical qubit with a specific error-correction scheme.
Specifically, an allocation statement is expressed as
\[
\tlogical\ q[m] := \tnew\ \<{\tt QEC}\>,
\]
where \(m\) logical codeblocks \(q[0],q[1],\ldots,q[m-1]\) are allocated, each consisting of multiple logical qubits. Here, \(\langle{\tt QEC}\rangle\) specifies the error-correction scheme associated with \(q\), ranging over the available error-correction schemes.

Conceptually, this statement is analogous to a classical declaration \(\mathbf{var}\ x := \tnew\ \mathtt{int}\), which allocates a variable \(x\) of type \(\mathtt{int}\).
This design enables more flexible resource utilization and more fine-grained resource estimation, as various error-correction schemes offer different trade-offs between resource consumption and error-protection capability, much like different types represent data with varying memory sizes.
It has been demonstrated that programmers can benefit from assigning error-correction schemes to functional modules according to their requirements; for example, distinct schemes are assigned to memory, processor, and resource modules to optimize overall resource consumption in the implementation of Shor's algorithm, as illustrated in Fig.~\ref{fig:shor}.

\paragraph{Programming Challenge: Intra- and Inter-Codeblock Operations}
Despite the flexibility and efficiency benefits, the association with specific error-correction schemes introduces challenges: logical operations can no longer be performed arbitrarily between any logical qubits, in contrast to fully abstracted models where all logical qubits are treated uniformly.
This restriction mirrors the classical setting, where operations between variables of different types are disallowed, thereby complicating the design of program statements and necessitating dedicated language support for logical operations, rather than directly lifting their physical counterparts as in existing languages.

In our language, primitive operations are divided into two categories: \emph{intra-codeblock} and \emph{inter-codeblock} operations.
The former require all operands to reside within the same codeblock, whereas the latter consist solely of \emph{joint Pauli measurements}, which are independent of the error-correction schemes associated with the participating codeblocks.
Joint Pauli measurements serve as a primitive building block for logical multi-qubit operations such as CNOT gates~\cite{1g44-jp62,Litinski2019gameofsurfacecodes}, and can be universally realized through \emph{code surgery} and \emph{adapters}~\cite{1g44-jp62}, which are conceptually analogous to type casting in classical programming by enabling operations across heterogeneous codeblocks.

\paragraph{Abstraction Challenge: Specification of Error-Correction Schemes}
As discussed above, multiple error-correction schemes may coexist, differing in their resource requirements, error-correction capabilities, and supported primitive operations.
This raises a key abstraction challenge: error-correction schemes must be carefully specified, rather than being simply treated as an error-suppression layer as in existing frameworks.

In our framework, specification of error-correction schemes is elevated from several backend parameters to a first-class abstraction that bridges the logical program and the underlying hardware configuration.
While the detailed specification will be presented later, we highlight here a distinctive feature of our framework: the explicit incorporation of classical decoding latency, which is crucial in fault-tolerant quantum computation~\cite{delfosse2023choosedecoderfaulttolerantquantum} but typically ignored by existing frameworks.
In particular, our model captures the trade-off between error suppression and decoding overhead: increasing the code distance generally reduces logical errors, but also increases decoding latency, potentially delaying measurement outcomes and allowing memory errors to accumulate.

\paragraph{Estimation Challenge: Cross-Layer Analysis of Resource Trade-offs}
The flexibility and resource efficiency enabled by the above design come at the cost of increased estimation complexity.
Since error-correction schemes encapsulate richer implementation-level information, resource estimation in our framework is influenced by a broader range of factors, such as the decoding latency mentioned above.
Consequently, unlike existing frameworks, where logical-level resource overheads can be estimated independently and subsequently translated into physical-level costs through an error-correction layer, our cross-layer design requires the logical program and its associated error-correction schemes to be analyzed jointly, thereby enabling more comprehensive and realistic resource estimation.

\begin{figure}[t]
    \centering
    \includegraphics[width=\textwidth]{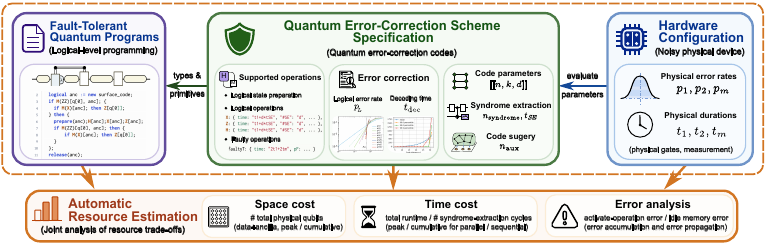}
    \caption{Overall architecture of our resource estimation framework for fault-tolerant quantum programs. }
    \label{fig:architecture}
\end{figure}

\paragraph{Organization and Contributions.}
Following the overall architecture shown in Fig.~\ref{fig:architecture}, we organize the remainder of the paper around the main components of our framework.
Sec.~\ref{sec:motex} first presents a motivating example of logical \(T\)-gate implementation, illustrating the programming support needed for practical fault-tolerant quantum computing and the benefits of exposing such support for resource utilization.
The technical contributions are then organized around four questions.
\begin{itemize}
    \item [Sec.~\ref{sec:prog}]
    \emph{How can a programming language express fault-tolerant quantum computing with heterogeneous QEC choices?}
    We introduce a language featuring codeblock allocation annotated with specific error-correction schemes, scheme-aware checking of intra-codeblock primitives, and joint Pauli measurements as the interface between codeblocks.
   \item [Sec.~\ref{sec:abstraction}]
\emph{How can underlying fault-tolerant support be abstracted to connect logical programs with physical resource estimation?}
We define a QEC/hardware interface that specifies program-visible information, such as supported primitives, together with estimator-required parameters, thereby bridging logical programs and physical resource estimates.

    \item [Sec.~\ref{sec:re}]
   \emph{How can resource trade-offs be analyzed compositionally over programs under fault-tolerant abstractions?}
We develop a conservative resource-relation inference system that follows the program structure and derives resource quantities from the underlying fault-tolerant abstractions, thereby enabling systematic exploration of time, space, and error trade-offs.

    \item [Sec.~\ref{sec:case}]
   \emph{Are the estimates consistent with established baselines, and what additional trade-offs do source-level QEC choices reveal?}
We implement a prototype and evaluate it through three case studies, showing three results, respectively: our framework supports finer-grained resource accounting and more effective resource utilization; it matches protocol-specific analyses; and it is broadly consistent with general tools such as QREv3~\cite{microsoft_qdk_qrev3}.
\end{itemize}
Finally, we discuss the impact and limitations of our framework in Sec.~\ref{sec:discussion} before reviewing related work in Sec.~\ref{sec:rw}.

\section{Motivating Example: Fault-Tolerant \(T\) Gate with Gate Teleportation}\label{sec:motex}
To illustrate the programming support required for practical fault-tolerant quantum computing, together with the resource-estimation process in our framework, we walk through a motivating example: the implementation of a logical \(T\) gate via gate teleportation, as shown between the processor and resource modules in Fig.~\ref{fig:shor}.

This process is typically treated as a black box characterized by several parameters in existing resource-estimation frameworks~\cite{microsoft_resource_estimator}, due to its inherent complexity. 
However, it is widely recognized that logical non-Clifford gates dominate the resource consumption of fault-tolerant quantum computing, and numerous magic-state distillation and injection protocols have been proposed~\cite{microsoft_t_factories,Litinski2019magicstate,Gidney2025magicstatecultivation_slides}. This highlights the importance of analyzing the resource costs of their concrete implementations in greater detail.

In this section, we focus on the gate-teleportation procedure. 
The complete resource-estimation results, together with a detailed implementation of magic-state distillation (corresponding to the resource module in Fig.~\ref{fig:shor}), are deferred to the case studies in Sec.~\ref{sec:case}.

\subsection{Implementation in Our Programming Language}
\begin{figure}[t]
    \centering
    \begin{minipage}{0.45\textwidth}
        \centering
        \includegraphics[scale=2]{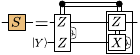}
        \caption{Implementation of a logical \(S\) gate using an ancilla in the state \(|Y\> := Y|+\>\).}
        \label{fig:logicalS}
        \vspace{2em}
        \includegraphics[scale=1.4]{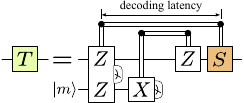}
        \caption{Implementation of a logical \(T\) gate using a magic state \(|m\>\) with Clifford \(S\)-correction.}
        \label{fig:logicalT}
    \end{minipage}
    \hspace{3em}
    \begin{minipage}{0.45\textwidth}
        \begin{lstlisting}[style=ftqp, basicstyle=\ttfamily\footnotesize]
def gateT(q: Codeblock) {
  logical anc := new surface_code;
//   call distill15to1(anc);
  if M(ZZ)[q[0], anc]; {
    // intermediate statement
    if M(X)[anc]; then Z[q[0]]; 
  } then {
    // implement logical S gate
    prepare(anc);
    H[anc];X[anc];Z[anc];
    if M(ZZ)[q[0], anc]; then {
        if M(X)[anc]; then Z[q[0]];
    }
  };
  release(anc);
}
        \end{lstlisting}
        \vspace{-0.7em}
        \caption{Source code for the logical \(T\) gate, where the call to the distillation protocol for preparing the magic state is temporarily commented out.}
        \label{fig:logicalT-code}
    \end{minipage}
\end{figure}

Unlike gates applied directly to physical qubits, a non-Pauli fault-tolerant logical gate is typically implemented through a protocol composed of multiple primitive operations.
As illustrated in Figs.~\ref{fig:logicalS} and~\ref{fig:logicalT}, the implementation of a logical \(S\) or \(T\) gate generally involves the following components:
(1) the participation of ancillary logical qubits prepared in specific states;
(2) the execution of logical joint Pauli measurements; and
(3) the application of conditional corrective operations based on the measurement outcomes.

\begin{table}[t]
  \centering
  \caption{Optimal error rates under different space budgets for the implementation of the logical \(T\) gate.}
  \label{tab:optimal-logicalT}
  \begin{tabular}{c c c c c}
  \toprule
  \multicolumn{1}{c}{Space budget}
  & \multicolumn{1}{c}{Optimal error rate}
  & \multicolumn{1}{c}{Error-correction schemes}
  & \multicolumn{1}{c}{Time cost} 
  & \multicolumn{1}{c}{Space cost} \\
  \midrule
  \(<4500\) qubits & 5.23\% & \((d_{\mathrm{data}}=27,d_{\mathrm{anc}}=27)\) & 2247 cycles & 4453 qubits\\
    \(<5000\) qubits & 4.31\% & \((d_{\mathrm{data}}=29,d_{\mathrm{anc}}=27)\) & 2769 cycles & 4789 qubits\\
    \(<5500\) qubits & 2.89\% & \((d_{\mathrm{data}}=31,d_{\mathrm{anc}}=29)\) & 3561 cycles & 5491 qubits\\
    \(<6000\) qubits & 2.40\% & \((d_{\mathrm{data}}=31,d_{\mathrm{anc}}=31)\) & 3741 cycles & 5857 qubits\\
  \bottomrule
  \end{tabular}
\end{table}
The source code for implementing the logical \(T\) gate in our language is shown in Fig.~\ref{fig:logicalT-code}.
In principle, operations whose implementations depend on the internal structure of the underlying error-correction schemes---such as logical Pauli gates---are treated as primitive operations in our language and are specified within the corresponding error-correction scheme.

Measurement-based conditional control flow is expressed using the \textbf{if} statement.
In contrast to physical measurements, the outcome of a logical measurement depends on the classical \emph{decoding} process, which introduces an inherent latency between the measurement and the execution of the corresponding controlled operations~\cite{delfosse2023choosedecoderfaulttolerantquantum}, as illustrated in Fig.~\ref{fig:logicalT-code}.

This latency is often neglected in existing frameworks; however, in practice, it typically exceeds the duration of a single primitive operation by a substantial margin, thereby impacting both the efficiency and the reliability of the program. Various techniques have been proposed to mitigate its effect, such as auto-corrected protocols~\cite{towards}, which will be discussed further in the case studies in Sec.~\ref{sec:case}. In our language, a middle statement can be executed between the measurement and the corresponding controlled operations, thereby enabling the effective utilization of the decoding latency.

\subsection{Resource Estimation}
Resource estimation in our framework focuses on three primary quantities: \emph{space cost} (the number of physical qubits), \emph{time cost}, and \emph{error rate}.
It is conducted jointly over the program, the specifications of the involved error-correction schemes (e.g., {\tt surface\_code} in Fig.~\ref{fig:logicalT-code}), and the configuration of the underlying hardware.
We omit a detailed discussion of the latter two components here. In brief, they enable the determination of the physical-qubit requirements, time cost, and logical error rate associated with a given codeblock and the operations it supports.

Under a conservative parameter setting based on current experimental reports, we assume a physical error rate of \(p_{\mathrm{phy}}=3\times 10^{-3}\) for superconducting qubits~\cite{google-nature}, a logical error rate per syndrome-extraction cycle of
\(
p_L = 0.1\left({p_{\mathrm{phy}}}/{0.57\%}\right)^{(d+1)/2}
\)
following~\cite{towards}, and a mean decoding latency estimated from Fig.~2(b) of~\cite{delfosse2023choosedecoderfaulttolerantquantum}. 
Based on these parameters, we estimate the resource costs of the program in Fig.~\ref{fig:logicalT-code}, using both data and ancillary qubits encoded with surface codes of distances ranging from \(d=27\) to \(d=31\).

The resource-estimation results automatically generated by our framework are summarized in Table~\ref{tab:optimal-logicalT}, where \(d_{\mathrm{data}}\) and \(d_{\mathrm{anc}}\) denote the code distances of the data and ancillary codeblocks (i.e., {\tt q} and {\tt anc} in Fig.~\ref{fig:logicalT-code}), respectively.
The results illustrate a clear trade-off among space cost, time cost, and error rate: 
by allocating more physical qubits to employ error-correction schemes with larger code distances, the overall error rate can be reduced, albeit at the expense of increased execution time.
Moreover, under a fixed space budget, the configuration obtaining the optimal error rate may involve assigning different code distances to data and ancillary codeblocks, as exemplified by the second and third rows of Table~\ref{tab:optimal-logicalT}, demonstrating the benefit of heterogeneous error-correction strategies for improving resource efficiency.

A more detailed investigation will be presented in Sec.~\ref{sec:case}, which further examines the asymmetric fault-tolerant requirements of data and ancillary qubits, as well as the corresponding optimal strategies for reducing error rates under limited resource constraints.


\section{A Fault-Tolerant Quantum Programming Language}\label{sec:prog}

In this section, we introduce a programming language for describing dynamic logical quantum circuits.
We first present its syntax and compilation scheme in Sec.~\ref{sec:language-syntax-semantics}, and then discuss its execution model with classical decoders in Sec.~\ref{sec:language-execution}.

\subsection{Syntax and Compilation}\label{sec:language-syntax-semantics}

The syntax of our programming language is defined in Fig.~\ref{fig:syntax}.
The language is written at the logical level: program variables are used to refer to logical qubits, while statements specify logical operations, measurements, control flow, and modular circuit structures.

Given a \emph{context} \(\Gamma\), defined as a list of distinct codeblocks, a statement \(S\) is compiled compositionally into a \emph{dynamic circuit} \(\sem{S}(\Gamma)\), i.e., a quantum circuit with classically controlled gates, provided that \(\ActBlock(S)\subseteq \Gamma\).
Here, \(\ActBlock(S)\) denotes the set of active codeblocks in \(S\), defined inductively in Fig.~\ref{fig:qubits}.
The context is a simplified version of the error context introduced later for resource estimation, in which each codeblock is additionally associated with an error-correction scheme and an error estimate.
The resulting circuit \(\sem{S}(\Gamma)\) has input and output interfaces exactly \(\Gamma\), as specified by the rules in Fig.~\ref{fig:compilation}.
We next explain the syntactic requirements of the language together with the corresponding compilation rule for each form of program statement.

\begin{figure}[t]
\[
\begin{aligned}
    &\mbox{(Pauli)} && P && ::= && I\mid X\mid Y\mid Z\mid P\otimes P\\
    &\mbox{(Measurement)} && \cM && ::= && \cM(P)\mid \cM(M)\\
    &\mbox{(Statement)} && S && ::= && \tlogical\ q[m]:=\tnew\ \<{\tt QEC}\>;S;\trelease(q)\\
    & && && \ \mid&& \tskip\mid \treset(q[i])\mid \mathrm{OP}[\bar{q}]\mid S;S\mid S\parallel S\\
    & && && \ \mid&& \tif\ \cM[\bar{q}];S_0\ \tthen\ S\mid
     \tcall\ \<{\tt name}\>(\bar{P},\bar{q})\\
    &\mbox{(Subcircuit)} && C && ::= && \tdef\ \<{\tt name}\>(\bar{P},\bar{q})\{S\}.
\end{aligned}
\]
\caption{Abstract syntax of the programming language.
Here, \(\langle{\tt QEC}\rangle\) denotes the name of an error-correction scheme specified later, \(\mathrm{OP}\) and \(M\) range over the primitive logical operations and measurements supported by the corresponding error-correction scheme, respectively, and \(\cM(P)\) denotes a joint Pauli measurement.}
\label{fig:syntax}
\end{figure}
\begin{figure}[t]
\[
\begin{aligned}
    &\ActBlock(\tlogical\ q[m]:=\tnew\ \<{\tt QEC}\>;S;\trelease(q))
    \triangleq\ActBlock(S)\setminus\{q[0],\ldots,q[m-1]\},\\
    &\ActBlock(\tskip)\triangleq\emptyset,\quad \ActBlock(\treset(q[i]))=\{q[i]\},\quad \ActBlock(\mathrm{OP}[\bar{q}])\triangleq\ActBlock(\bar{q}),\\
    &\ActBlock(S_1;S_2)\triangleq\ActBlock(S_1)\cup \ActBlock(S_2),\\
    &\ActBlock(S_1\parallel S_2)\triangleq \ActBlock(S_1)\cup \ActBlock(S_2),\\
    &\ActBlock(\tif\ \cM[\bar{q}];S_0\ \tthen\ S)\triangleq\ActBlock(S_0)\cup \ActBlock(S)\cup \ActBlock(\bar{q}),\\
    &\ActBlock(\tcall\ \<{\tt name}\>(\bar{P},\bar{q}))\triangleq\ActBlock(\bar{q}).
\end{aligned}
\]
\caption{Definition of \(\ActBlock(S)\), where \(\ActBlock(\bar{q})\triangleq\{q[i]\mid\exists j, q[i][j]\in \bar{q}\}\) extracts the codeblocks accessed in \(\bar{q}\).}
\label{fig:qubits}
\end{figure}
\begin{figure}[t]
    \centering
    \begin{minipage}{0.3\textwidth}
        \centering
        \includegraphics[scale=1.2]{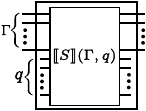}
        \subcaption{\(\sem{\tlogical\ q[m]\ldots\<{\tt QEC}\>}(\Gamma)\)}
        \label{fig:compilation-1}
    \end{minipage}
    \hspace{0.5em}
    \begin{minipage}{0.25\textwidth}
        \centering
        \includegraphics[scale=1.2]{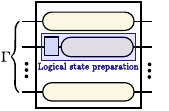}
        \subcaption{\(\sem{\treset(q[i])}(\Gamma)\)}
        \label{fig:compilation-2}
    \end{minipage}
    \begin{minipage}{0.4\textwidth}
        \centering
        \includegraphics[scale=1.2]{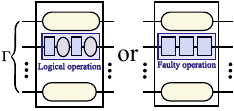}
        \subcaption{\(\sem{\mathrm{OP}[\bar{q}]}(\Gamma)\)}
        \label{fig:compilation-3}
    \end{minipage}
    \\[0.5em]
    \begin{minipage}{0.45\textwidth}
        \centering
        \includegraphics[scale=1.2]{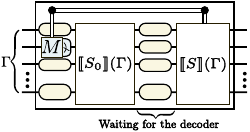}
        \subcaption{\(\sem{\tif\ \cM(M)[\bar{q}];S_0\ \tthen\ S}(\Gamma)\)}
        \label{fig:compilation-5}
    \end{minipage}
    \begin{minipage}{0.45\textwidth}
        \centering
        \includegraphics[scale=1.2]{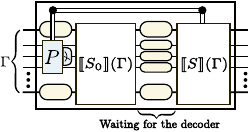}
        \subcaption{\(\sem{\tif\ \cM(P)[\bar{q}];S_0\ \tthen\ S}(\Gamma)\)}
        \label{fig:compilation-4}
    \end{minipage}\\[0.5em]
    \begin{minipage}{0.3\textwidth}
        \includegraphics[scale=1.2]{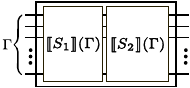}
        \subcaption{\(\sem{S_1;S_2}(\Gamma)\)}
        \label{fig:compilation-6}
    \end{minipage}
    \hspace{.5em}
    \begin{minipage}{0.25\textwidth}
        \includegraphics[scale=1.2]{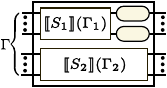}
        \subcaption{\(\sem{S_1\parallel S_2}(\Gamma)\)}
        \label{fig:compilation-7}
    \end{minipage}
    \hspace{.5em}
    \begin{minipage}{0.4\textwidth}
        \hspace{1em}
        \includegraphics[scale=1.2]{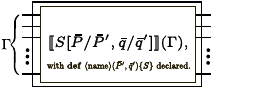}
        \subcaption{\(\sem{\tcall\ \<{\tt name}\>(\bar{P},\bar{q})}(\Gamma)\)}
        \label{fig:compilation-8}
    \end{minipage}
    \hspace{-3em}
    \caption{Compilation of program statements into dynamic fault-tolerant circuits.
    Each horizontal wire denotes one codeblock.
    Yellow boxes (\protect\yellowbox{}) denote repeated syndrome-extraction cycles;
    light blue boxes (\,\protect\bluebox{}\,) represent physical operations and measurements acting on physical qubits;
    and dark blue boxes (\,\protect\dbluebox{}\,) mark components whose internal physical implementation is abstracted away and treated as a black box by the programming language, with cost and error behavior specified later in the error-correction scheme.
    }
    \label{fig:compilation}
\end{figure}

\paragraph{Addressing and allocation of codeblocks.}
The statement \(\tlogical\ q[m] := \tnew\ \<{\tt QEC}\>\) allocates \(m\) codeblocks associated with the error-correction scheme \(\<{\tt QEC}\>\), which is specified at the meta-program level discussed in Sec.~\ref{sec:abstraction}.
If the scheme has parameters \([[n,k,d]]\), then each codeblock \(q[i]\) contains \(k\) logical qubits, denoted by \(q[i][0],\ldots,q[i][k-1]\).
Fig.~\ref{fig:compilation-1} illustrates the compilation of this allocation statement.
The body \(S\) is compiled under the extended context \(\Gamma,q[0],\ldots,q[m-1]\); upon completion of \(S\), these local codeblocks are released, restoring the external interface to \(\Gamma\).

\paragraph{Logical state preparation and primitive operations.}
The statement \(\treset(q[i])\) prepares the codeblock \(q[i]\) in a prescribed state, whose concrete form is abstracted away since our focus is resource estimation.
The statements \(\mathrm{OP}[\bar{q}]\) and \(\cM(M)[\bar{q}]\) denote intra-codeblock primitive operations and measurements, respectively.
They are well-formed only when all operands in \(\bar{q}\) belong to the same codeblock, i.e., \(\bar{q}=q[i][j_1],\ldots,q[i][j_a]\), and the associated error-correction scheme supports the operation \(\mathrm{OP}\) or measurement \(M\) with arity \(a\).
A primitive may be either fault-tolerant or faulty, as specified by the error-correction scheme.
Faulty primitives model uncorrected physical operations applied directly to physical qubits, without intervening syndrome extraction or error suppression.
They are typically cheaper and easier to implement, such as a noisy physical \(T\) gate, but induce relatively high logical-level error rates.
Nevertheless, such primitives can be exploited by distillation protocols to produce high-fidelity outputs, as discussed later.

Figs.~\ref{fig:compilation-2} and~\ref{fig:compilation-3} illustrate the compilation of preparation and primitive statements, which are treated as black-box primitives at the programming-language level and specified abstractly by the error-correction scheme.
Logical state preparation typically consists of physical reset followed by syndrome-extraction cycles.
Primitive logical operations combine physical operations with interleaved syndrome extraction, whereas primitive faulty operations consist only of physical operations.
To reflect this abstraction, these components are enclosed in dark-blue boxes (\,\dbluebox\,), indicating that their internal implementations are hidden.

During the execution of a primitive operation, only the involved codeblock is active, while all other codeblocks remain idle and are therefore subject to memory errors\footnote{At the logical level, we refer to these as \emph{memory errors}, arising from quantum \emph{decoherence}, which limits the ability of physical qubits to preserve high-fidelity logical states over time.}.
To mitigate such errors, periodic syndrome extraction is performed to enable classical decoding and error correction.

\begin{figure}[t]
    \centering
    \includegraphics[scale=1.2]{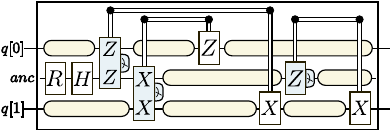}
    \caption{Compiled circuit for the CNOT gate. Adjacent syndrome-extraction intervals are merged into longer yellow boxes. The symbol \(R\) denotes logical state preparation, and \(H\), \(X\), and \(Z\) denote primitive logical gates.}
    \label{fig:CNOT}
\end{figure}
\paragraph{Measurement-based conditionals.}
The statement \(\tif\ \cM[\bar{q}];S_0\ \tthen\ S\) denotes a measurement-based conditional, where the execution of \(S\) depends on the measurement outcome.
In addition to the primitive measurements discussed above, our language supports joint Pauli measurements \(\cM(P)\) across codeblocks, allowing the target qubits to reside in different codeblocks.
Since the measurement outcome may become available only after classical decoding, an intermediate statement \(S_0\) is inserted to execute during the decoding latency.

Figs.~\ref{fig:compilation-5} and~\ref{fig:compilation-4} illustrate the compilation of measurement-based conditionals for primitive measurements and joint Pauli measurements, respectively.
In both cases, the statement \(S\) is executed only after the measurement outcome has become available and the intermediate statement \(S_0\) has completed; additional syndrome-extraction cycles are inserted whenever the decoding latency exceeds the execution time of \(S_0\).

\paragraph{Composition and modularity.}
Both sequential and parallel compositions are supported in our language, with their compilation rules shown in Figs.~\ref{fig:compilation-6} and~\ref{fig:compilation-7}, respectively.
For parallel composition \(S_1\parallel S_2\), the two statements are required to act on disjoint sets of codeblocks, i.e., \(\ActBlock(S_1)\cap \ActBlock(S_2)=\emptyset\).
To compile \(S_1\parallel S_2\), the context is partitioned into two disjoint subcontexts
\(
  \Gamma = \Gamma_1 \uplus \Gamma_2
\)
such that \(\ActBlock(S_1)\subseteq \Gamma_1\) and \(\ActBlock(S_2)\subseteq \Gamma_2\).
The resulting circuit \(\sem{S_1\parallel S_2}(\Gamma)\) is then obtained by juxtaposing the two subcircuits \(\sem{S_1}(\Gamma_1)\) and \(\sem{S_2}(\Gamma_2)\).
This construction is well-defined for any such partition, since \(\sem{S}(\Gamma)\) acts on the blocks in \(\Gamma\setminus \ActBlock(S)\) only through syndrome extraction. If the two branches have different durations, additional syndrome-extraction cycles are appended to the shorter branch to synchronize their execution.

To support modular programming, our language provides subcircuit declarations and invocations with formal parameters for both Pauli operators and codeblocks.
As shown in Fig.~\ref{fig:compilation-8}, a subcircuit call is compiled by substituting the actual parameters for the corresponding formal parameters in the subcircuit body.
\begin{example}\label{eg:CNOT}
Suppose that \({\tt surface\_code}\) specifies a single-logical-qubit codeblock supporting logical \(H\), \(X\), and \(Z\) gates.
A logical \(\rCNOT\) between two encoded qubits can be expressed by allocating one ancillary block and using joint Pauli measurements:
\[
\begin{aligned}
\rCNOT[q[0],q[1]] \equiv\;&
\tlogical\ anc:=\tnew\ {\tt surface\_code};\\
&\treset(anc);H[anc];\\
&\tif\ \cM(ZZ)[q[0],anc];
   (\tif\ \cM(XX)[anc,q[1]];\tskip\ \tthen\ Z[q[0]])\\
&\qquad \tthen\ X[q[1]];\\
&\tif\ \cM(Z)[anc];\tskip\ \tthen\ X[q[1]];\\
&\trelease(anc).
\end{aligned}
\]
Under the compilation rules above, this program yields the dynamic circuit in Fig.~\ref{fig:CNOT}, where adjacent syndrome-extraction intervals are merged.

\end{example}

\subsection{Execution with Classical Decoders}\label{sec:language-execution}
After obtaining a circuit consisting of logical operations and syndrome extractions, it is necessary to address how the circuit can be executed in a fault-tolerant manner.
This typically relies on a \emph{classical decoder}, which processes the classical information extracted from syndrome measurements to perform error correction.

Such an execution model is crucial for subsequent resource estimation, as the execution of the circuit depends on classical feedback, and thus the classical overhead must be taken into account.
In modern fault-tolerant quantum computing, however, standard Pauli frame tracking techniques~\cite{PennyLane-PauliFrameTracking} can effectively delay and merge Pauli operations, thereby decoupling quantum circuit execution from the classical decoding process.
This significantly simplifies the execution model and leads to more streamlined resource estimation.

\begin{figure}[t]
    \centering
    \includegraphics[scale=1.9]{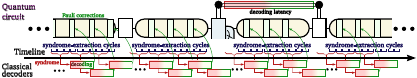}
    \caption{Execution of syndrome extraction with classical decoding.
    A sliding window of syndrome data is sent to a decoder, represented by a gradient-colored box (\protect\gradbox{}).
    Red arrows indicate syndrome inputs to the decoder, and green arrows indicate the resulting Pauli-frame updates or corrections.}
    \label{fig:decoding}
\end{figure}

Fig.~\ref{fig:decoding} illustrates a sliding-window decoding model~\cite{delfosse2023choosedecoderfaulttolerantquantum,dennis2002topological}, in which syndrome extraction proceeds cycle by cycle and, after a fixed window of cycles is accumulated, the corresponding syndrome data are transmitted to a classical decoder, whose output specifies the Pauli corrections to be applied.
In practice, however, physical Pauli corrections need not be executed immediately. Instead, standard Pauli frame tracking~\cite{PennyLane-PauliFrameTracking} records these corrections classically and updates the interpretation of subsequent measurements accordingly, thereby eliminating the need to stall the quantum device for each decoder output.
A more detailed discussion about this process is provided in the appendix.

Classical decoding remains essential for measurement-based control, since the effective outcome of a logical measurement is determined only after the associated syndrome data have been decoded.
Consequently, in the conditional statement \(\tif\ \cM[\bar{q}];S_0\ \tthen\ S\), the guarded statement \(S\) may need to wait for both the completion of \(S_0\) and the availability of the decoded outcome, during which idle codeblocks continue to accumulate memory errors.

Finally, decoding imposes its own throughput constraints: if syndrome data are produced faster than the available classical resources can process them, a \emph{backlog} problem may arise~\cite{backlog}.
In this work, we do not model decoder hardware in detail; instead, we assume sufficient parallel classical capacity as in Fig.~\ref{fig:decoding} to sustain steady-state syndrome extraction and account explicitly only for the latency that impacts measurement-based control.


\section{Abstraction of Fault-Tolerant Execution Platforms}\label{sec:abstraction}
To perform resource estimation for logical-level circuits expressed in the programming language introduced above, in this section we abstract the underlying fault-tolerant execution platform, including both the \emph{specifications of quantum error-correction schemes} and the \emph{hardware configurations}, as illustrated in Fig.~\ref{fig:architecture}.

\subsection{Configuration of Noisy Physical Hardware}
The configuration of noisy physical hardware follows the same coarse-grained
abstraction as Microsoft's Quantum Resource Estimator~\cite{microsoft_resource_estimator}, where all physical qubits are assumed to be homogeneous. Their properties are
characterized along two dimensions: physical error rates and physical operation
durations.

As illustrated in Fig.~\ref{fig:architecture}, a hardware configuration \(\cC\) is
formally defined as the tuple
\[
\cC=(p_1,p_2,p_m,t_1,t_2,t_m),
\]
where \(p_1\), \(p_2\), and \(p_m\) denote the error rates of single-qubit gates, two-qubit gates, and single-qubit measurements (readout), respectively, and \(t_1\), \(t_2\), and \(t_m\) denote their corresponding execution times.
\begin{example}\label{eg:config-sc}
  As a representative running example for the subsequent discussion, we consider the following superconducting-qubit configuration with conservatively chosen parameters:
\[
\cC_{\mathrm{sc}}
=
\pare{
    p_1=10^{-3},\,
    p_2=2\times 10^{-3},\,
    p_m=3\times 10^{-3},\,
    t_1=10^{-8}\,\mathrm{s},\,
    t_2=10^{-7}\,\mathrm{s},\,
    t_m=10^{-6}\,\mathrm{s}
}.
\]
These parameters correspond to nanosecond-scale gate operations, microsecond-scale measurements, and physical error rates on the order of \(10^{-3}\), which are broadly consistent with current experimental reports for superconducting qubits~\cite{google-nature}.
\end{example}


\subsection{Specification of Quantum Error-Correction Schemes}\label{sec:spec}

\begin{table}[t]
  \centering
  \caption{Description of the fields used in an error-correction scheme specification.}
  \label{table:qecc-specification}
  \renewcommand{\arraystretch}{1.3}
  \begin{tabularx}{\textwidth}{c X}
    \toprule
    \textbf{Field} & \multicolumn{1}{c}{\textbf{Brief description}} \\
    \midrule
    \([[n,k,d]]\)
      & Basic code parameters: each codeblock encodes \(k\) logical qubits using \(n\) physical qubits with code distance \(d\). \\
    \(\mathrm{SE}\)
      & Syndrome-extraction procedure: the number of physical syndrome qubits and the time required for one syndrome-extraction cycle. \\
    \(\mathrm{Prep}\)
      & Logical state preparation: the required time, number of syndrome-extraction cycles, and ancillary qubits. \\
    \(\mathrm{OP}_L\)
      & A list of supported primitive logical operations, specifying the required time, number of syndrome-extraction cycles, and ancillary qubits for each operation. \\
    \(\mathrm{OP}_F\)
      & A list of supported primitive faulty operations, specifying the error rate, required time, and ancillary qubits for each operation. \\
    \(t_{\mathrm{dec}}\)
      & Decoding-time estimate for obtaining logical measurement outcomes. \\
    \(p_L\)
      & Logical error rate per syndrome-extraction cycle. \\
    \(n_{\mathrm{aux}}\)
      & Number of ancillary qubits required for auxiliary graph surgery. \\
    \bottomrule
  \end{tabularx}
\end{table}



The specification of quantum error-correction schemes serves as a crucial bridge between the logical-level programs and the noisy physical hardware.
In our framework, a quantum error-correction scheme named \(\<{\tt QEC}\>\) is formally specified as a tuple
\[
\cS_{\<\tt QEC\>}=\pare{[[n,k,d]],\,\mathrm{SE},\,\mathrm{Prep},\,\mathrm{OP}_L,\,\mathrm{OP}_F\,,t_{\mathrm{dec}},\,p_L,\,n_{\mathrm{aux}}},
\]
where Table~\ref{table:qecc-specification} briefly describes each field; a more detailed explanation is given below.


\paragraph{Syndrome extraction, state preparation and primitive operations.}
The entry \(\mathrm{SE}=(n_{\mathrm{syn}},t_{\mathrm{SE}})\) consists of an expression \(n_{\mathrm{syn}}\) over the parameters \(n,k,d\), which records the number of ancillary qubits required for syndrome extraction, and an expression \(t_{\mathrm{SE}}\) over the parameters \(n,k,d,t_1,t_2,t_m\) configured in the underlying \(\cC\), which records the time cost of one syndrome-extraction cycle.

\begin{example}\label{eg:surface-spec}
Suppose an \([[n,k,d]]\) surface code requires, in terms of circuit depth, two single-qubit gates, four two-qubit gates, and one measurement to complete a single syndrome-extraction cycle. The corresponding syndrome-extraction specification can then be written as
\[
\mathrm{SE}=(n_{\mathrm{syn}}=n-1,\; t_{\mathrm{SE}}=2t_1+4t_2+t_m)\in \cS_{\tt surface\_code}.
\]
For the instance \(n=225, k=1, d=15\), together with the hardware configuration \(\cC_{\mathrm{sc}}\) in Example~\ref{eg:config-sc}, these expressions yield the following concrete estimates for the syndrome-extraction cost:
\[
\sem{n_{\mathrm{syn}}}_{\cS_{\tt surface\_code},\cC_{\mathrm{sc}}}=224,
\qquad
\sem{t_{\mathrm{SE}}}_{\cS_{{\tt surface\_code}},\cC_{\mathrm{sc}}}
=2\times 10^{-8}+4\times 10^{-7}+10^{-6}
=1.42\times 10^{-6}.
\]
In the sequel, we abbreviate \(\sem{e}_{\cS,\cC}\) as \(\sem{e}_{\cS}\) when the hardware configuration \(\cC\) is clear from context.
\end{example}

The entry \(\mathrm{Prep}=\pare{t_{\mathrm{prep}},n_{\mathrm{SE}},n_{\mathrm{anc}}}\) specifies the time cost, number of syndrome-extraction cycles, and ancillary qubit requirement for preparing the initial logical state.
The entry \(\mathrm{OP}_L=[L_1,L_2,\ldots,L_m]\) denotes the collection of supported primitive logical operations with fault-tolerant implementations.
A logical operation specification \(L\) is defined as a tuple
\[
L=\pare{\text{arity}_L,t_L,n_{\mathrm{SE}},n_{\text{anc}}},
\]
where \(\mathrm{arity}_L\) is the operation arity, while \(t_L\), \(n_{\mathrm{SE}}\), and \(n_{\mathrm{anc}}\) specify the time cost, required syndrome-extraction cycles, and ancillary qubits, respectively.

\begin{example}
A logical \(X\) gate on a \(d=15\) surface-code patch can be implemented by applying 15 physical \(X\) gates in parallel, followed by \(d=15\) cycles of syndrome extraction. This operation can be specified as
\[
X=(1,t_1+d\cdot t_{\mathrm{SE}}, d, 0),
\]
where \(d\) and \(t_{\mathrm{SE}}\) are parameters introduced previously.
\end{example}

The specification of faulty operations is given by a list \(\mathrm{OP}_F = [F_1, F_2, \ldots, F_m]\), where each faulty operation specification \(F\) is defined as a tuple
\[
F = \pare{1, t_F, p_F, n_{\mathrm{anc}}},
\]
which is similar to the specification of logical operations, except that the error rate \(p_F\) is given explicitly as a parameter.
This is because no syndrome extraction or error correction is performed during the execution of a faulty operation, and hence its error rate is not suppressed by the error-correction scheme.

\paragraph{Logical error rate and decoding time.}
The entry \(p_L\) denotes the logical error rate per syndrome-extraction cycle, which typically depends on both the code distance and the underlying physical error rate.
For example, given a hardware configuration, one may either directly specify \(p_L = 1.1\times 10^{-5}\) as a constant obtained from experimental characterization, or alternatively specify \(p_L\) as an expression using a commonly adopted heuristic formula parameterized by the code distance \(d\) and physical error rate \(p\)~\cite{towards, microsoft_resource_estimator, delfosse2023choosedecoderfaulttolerantquantum, assessing,google-nature}:
\[
p_L = a\cdot(p/p_{\mathrm{th}})^{(d+1)/2}\mbox{ with chosen constants }a\mbox{ and }p_{\mathrm{th}}.
\]

The entry \(t_{\mathrm{dec}}\) denotes an estimate of the classical decoding time, which is required for timing analyses of measurement-based conditional statements, as discussed in the previous section.
The decoding time is typically modeled as a random variable with distribution \(\Pr_{\mathrm{dec}}(t)\), which depends on the physical error rate \(p\) and the code distance \(d\), and represents the probability that the decoding procedure completes in time \(t\)~\cite{delfosse2023choosedecoderfaulttolerantquantum}.

Depending on the desired level of conservativeness, different estimates of the decoding time may be used, ranging from the worst-case decoding time
\(
t_{\mathrm{dec}} := \sup \{ t : \Pr_{\mathrm{dec}}(t) > 0 \}
\)~\cite{delfosse2023choosedecoderfaulttolerantquantum},
to a high-probability bound given by the 99.9th percentile
\(
t_{\mathrm{dec}} := \inf \{ t : \sum_{t' \le t} \Pr_{\mathrm{dec}}(t') \ge 0.999\}.
\)
Both choices can be represented either as constants or as expressions derived from the parameters introduced above.
For example, from the experimental results shown in Fig.~2(b) of~\cite{delfosse2023choosedecoderfaulttolerantquantum}, the 99.9th percentile decoding time for a \(d=27\) surface code is approximately \(255\) cycles of syndrome extraction. This can be expressed as
\(
t_{\mathrm{dec}} = 255\cdot t_{\mathrm{SE}},
\)
where \(t_{\mathrm{SE}}\) denotes the duration of one syndrome-extraction cycle as introduced previously.

\paragraph{Ancillary qubits for auxiliary graph surgery.}
The entry \(n_{\mathrm{aux}}\) denotes the number of ancillary qubits required for auxiliary graph surgery, which provides the interface for universal adapters enabling joint Pauli measurements across heterogeneous codeblocks~\cite{1g44-jp62}.
Intuitively, auxiliary graph surgery introduces an auxiliary graph associated with each codeblock, where edges correspond to ancillary data qubits, and vertices and faces correspond to ancillary syndrome qubits.
The resource cost of such a construction is determined by the underlying error-correction scheme; for example, it requires \(O(d\log^3 d)\) ancillary qubits for general quantum LDPC codes and \(O(d)\) for codes with special geometrically local structure~\cite{1g44-jp62}.
Accordingly, \(n_{\mathrm{aux}}\) is used to specify the number of ancillary qubits required for auxiliary graph surgery under the given error-correction scheme.

\begin{figure}[t]
    \begin{lstlisting}[style=ftqp, basicstyle=\ttfamily\footnotesize]
qecc surface_code using superconducting = {
  [[n,k,d]]: [729, 1, 27],
  SE: { "#ancilla": "n-1", tSE: "2t1+4t2+tm" },
  PREP: { time: "t1+d*tSE", "#SE": "d", "#ancilla": "0" },
  OPL: {
    I: { arity: 1, time: "0",        "#SE": "0", "#ancilla": "0" },
    X: { arity: 1, time: "t1+d*tSE", "#SE": "d", "#ancilla": "0" },
    Z: { arity: 1, time: "t1+d*tSE", "#SE": "d", "#ancilla": "0" },
    H: { arity: 1, time: "t1+d*tSE", "#SE": "d", "#ancilla": "0" },
  },
  OPF: { faultyT: { arity: 1, time: "2t1+2tm", pF: "max(p1,p2,pm)/30", "#ancilla": "0" } },
  tdec: "255*tSE",
  pL: "0.1*(max(p1,p2,pm)/0.0057)^((d+1)/2)",
  naux: "d*d"
};
    \end{lstlisting}
    \caption{An example of an error-correction scheme specification for the surface code with \(d=27\), written in the format used by our prototype implementation.
  Here, we use a relatively conservative error threshold \(p_{\mathrm{th}}=0.57\%\), following the numerical simulations of surface-code logical error rates reported in Fig.~4 of~\cite{towards}.
  For a physical error rate \(p\approx 3\times 10^{-3}\), this gives an
  error-suppression factor
  \(
  \Lambda \approx {p_{\mathrm{th}}}/{p} \approx 2,
  \)
  which is consistent with the experimental results reported in~\cite{google-nature}.
  The number of ancillary qubits required for auxiliary graph surgery is estimated as \(n_{\mathrm{aux}} = d^2\), since lattice surgery is a special case of auxiliary graph surgery~\cite{1g44-jp62}.
  }
    \label{fig:qecc-specification}
\end{figure}

Finally, an example of the error-correction scheme specification for the surface code (\(d=27\)) is shown in Fig.~\ref{fig:qecc-specification}, where {\tt X} denotes the logical \(X\) gate and {\tt faultyT} denotes the faulty\(T\) measurement operation, which will be illustrated in case studies in Sec.~\ref{sec:case}.

\section{Resource Estimation and Error Analysis}\label{sec:re}

In this section, we present an inference system for compositionally estimating the resource requirements of fault-tolerant quantum programs, building on the programming language and abstractions introduced above.
The two primary resource metrics considered in quantum computing, and in our framework, are \emph{time cost} and \emph{space cost}, where the latter refers to the number of physical qubits.

In addition, we analyze the \emph{error rate} of each logical codeblock in fault-tolerant quantum programs, due to its inherent trade-off with resource cost:
longer execution times lead to additional memory errors, while stricter error requirements may increase the resource cost.
%
Essentially, we are concerned with reasoning about a big-step resource relation of the following form:
\begin{equation}\label{eq:re-semantics}
(S,\Gamma)\Downarrow(\Gamma',T,N),
\end{equation}
where \(S\) is a program statement, \(\Gamma\) and \(\Gamma'\) are error contexts to be defined later, and \(T\in\bR_{\geq 0}\) and \(N\in\bN\) are constants.
Intuitively, this relation states that executing the program statement \(S\) in the error context \(\Gamma\) produces a new error context \(\Gamma'\), incurs time cost \(T\), and requires \(N\) auxiliary physical qubits beyond those already contained in the codeblocks of \(\Gamma\).
Together, these components capture both the resource consumption and the error behavior of \(S\).

An \emph{error context} \(\Gamma\) records, for each live codeblock, the error-correction scheme associated with that block and the current probability that the block has been disturbed by errors.
We track error rates separately for each codeblock, since most operations act only on a small local set of codeblocks and therefore do not introduce errors to the rest of the system.
As illustrated later, our analysis also accounts for error propagation induced by joint operations and measurement-based control flow.
Formally, we define
\[
\Gamma::=\bullet \mid \Gamma, Q :_{\epsilon} \cS,
\]
where \(Q :_{\epsilon} \cS\) denotes that the codeblock \(Q\) is associated with the quantum error-correction scheme \(\cS\).
The parameter \(\epsilon\in[0,1]\) represents the probability that the codeblock \(Q\) has been disturbed by errors in the current context, or equivalently, the codeblock remains ideal with probability \(1-\epsilon\).
An error context \(\Gamma\) is also viewed as a finite map and for a codeblock \(Q\) we write
\[
\Gamma(Q)=\pare{\epsilon^\Gamma_Q,\cS^\Gamma_Q},
\]
where \(\epsilon^\Gamma_Q\) and \(\cS^\Gamma_Q\) denote the error probability and error-correction scheme associated with \(Q\), respectively.
We use \(\dom(\Gamma)\) to denote the set of codeblocks in \(\Gamma\).


Two auxiliary notations are used throughout this section.
First, if a codeblock has failed with probability \(\epsilon\), and an additional independent failure occurs with probability \(p\), we write the combined failure probability as
\[
\epsilon\oplus p := 1-(1-\epsilon)(1-p).
\]
Second, for a codeblock \(Q\in\dom(\Gamma)\) and a time interval \(T\), we define the accumulated idle-memory error estimate as
\begin{equation}\label{eq:idle}
\mathsf{Idle}_{\Gamma}(Q,T)
:=
\left\lceil
\frac{T}{\sem{t_{\mathrm{SE}}}_{\cS^\Gamma_Q}}
\right\rceil
\cdot
\sem{p_L}_{\cS^\Gamma_Q}
\in\bR_{\geq 0}.
\end{equation}
Here, \(t_{\mathrm{SE}}\) and \(p_L\) denote the time cost of one syndrome-extraction cycle and the logical error rate per syndrome-extraction cycle specified by \(\cS^\Gamma_Q\), respectively, as introduced in Sec.~\ref{sec:spec}.
Following Example~\ref{eg:surface-spec}, we write \(\sem{\cdot}_{\cS}\) for the evaluation of an expression under the error-correction scheme \(\cS\), omitting the hardware configuration \(\cC\) when it is clear from the context.

The estimate in Eq.~\eqref{eq:idle} is obtained by multiplying the logical error rate per syndrome-extraction cycle by the number of cycles incurred during the idle period. The latter is approximated by the ratio between the elapsed time \(T\) and the duration of a single syndrome-extraction cycle, since idle codeblocks are assumed to continuously undergo syndrome extraction throughout execution, as discussed in Sec.~\ref{sec:prog}.

This approach yields a coarse yet conservative approximation: instead of analyzing detailed fault configurations within individual cycles, it aggregates the per-cycle logical error probabilities.
Nevertheless, when the logical error rate is sufficiently small, the approximation is first-order accurate and consistent with experimental observations; see, for example, the approximately linear behavior reported in Fig.~2d of~\cite{google-nature}.

It is also convenient to package the common context update into a single notation.
Given a set of active codeblocks \(\cB\subseteq\dom(\Gamma)\), an updated failure estimate \(\epsilon\), and an elapsed time \(T\), let \(\mathsf{Step}_{\Gamma}(\cB,\epsilon,T)\) be the context with domain \(\dom(\Gamma)\), defined for each \(Q\in\dom(\Gamma)\) by
\begin{equation}\label{eq:step}
\mathsf{Step}_{\Gamma}(\cB,\epsilon,T)(Q)
=
\begin{cases}
\pare{\epsilon,\cS^\Gamma_Q},
& \text{if } Q\in\cB,\\[0.8em]
\pare{\epsilon^\Gamma_Q\oplus \mathsf{Idle}_{\Gamma}(Q,T),\cS^\Gamma_Q},
& \text{otherwise}.
\end{cases}
\end{equation}
Intuitively, \(\mathsf{Step}_{\Gamma}(\cB,\epsilon,T)\) updates the error context after an execution step of duration \(T\): the error rates of active codeblocks in \(\cB\) are updated to \(\epsilon\), while all other codeblocks accumulate idle-memory errors over the same time interval.
For a pure waiting step, we write
\[
\mathsf{Wait}_{\Gamma}(T)
:=
\mathsf{Step}_{\Gamma}(\emptyset,0,T).
\]
In addition, to facilitate counting the ancillary qubits required for syndrome extraction in currently idle codeblocks, we define, for \(\cB \subseteq \dom(\Gamma)\), the number of syndrome qubits required for all codeblocks in \(\cB\) as
\[
\mathsf{AncSE}_\Gamma(\cB)=\sum_{Q\in\cB}\sem{n_{\mathrm{syn}}^{(Q)}}_{\cS^\Gamma_Q},
\]
where \(n_{\mathrm{syn}}^{(Q)}\) denotes the number of physical syndrome qubits required for a single syndrome-extraction cycle specified by \(\cS^\Gamma_Q\) for the codeblock \(Q\).

One major advantage of a programming language is that it allows us to account for resource consumption in a structured and compositional way.
Therefore, we next discuss the reasoning rules for the resource relations case by case, following the structure of the program statement \(S\).

\subsection{Primitive Operations, State Preparation and Codeblock Allocation}\label{sec:primitive-operation}
For a primitive logical operation statement \(L[q[i][j_1],\ldots,q[i][j_a]]\), we have the following inference rule:
{  \small
\[
\inferrule{
    \cS:=\cS^\Gamma_{q[i]}\\
    L=(a,t,n_{\mathrm{SE}},n_{\mathrm{anc},L})\in \mathrm{OP}_L\in\cS\\
    \mathrm{SE}=(n_{\mathrm{syn}},t_{\mathrm{SE}})\in\cS\\
    p_L\in \cS\\
    \epsilon:=\epsilon^{\Gamma}_{q[i]}\oplus\sem{n_{\mathrm{SE}}}_{\cS}\sem{p_L}_{\cS}\\
    T:=\sem{t}_{\cS}\\
    N:=\sem{n_{\mathrm{anc},L}}_{\cS}+\mathsf{AncSE}_\Gamma(\dom(\Gamma))
}{
    \pare{L[q[i][j_1],\ldots,q[i][j_a]],\Gamma}\Downarrow\pare{\textsf{Step}_{\Gamma}(\{q[i]\},\epsilon,T),T,N}
}{}.
\]
}

\noindent The time cost \(T\) and space cost \(N\) of a primitive logical operation are obtained directly from the associated error-correction scheme.
For a logical operation \(L\) on codeblock \(q[i]\), let \(\cS:=\cS^\Gamma_{q[i]}\) be its associated error-correction specification.
The time cost is given by evaluating the expression \(t\) specified for \(L\) under \(\cS\), while the additional space cost is the sum of the ancillary qubits required by \(L\) itself and those required for syndrome extraction of all codeblocks.

The resulting error context is \(\mathsf{Step}_\Gamma(\{q[i]\},\epsilon,T)\), where \(q[i]\) is the only active codeblock.
The post-operation failure probability is estimated as
\(
\epsilon^{\Gamma}_{q[i]} \oplus
\sem{n_{\mathrm{SE}}}_{\cS}
\cdot
\sem{p_L}_{\cS}
\),
which captures the probability that the codeblock \(q[i]\) has already been affected by errors prior to the operation, or that the operation itself fails.
The latter adopts the same accumulated-error approximation as in the definition of \(\mathsf{Idle}\) in Eq.~\eqref{eq:idle}, namely the product of the number of syndrome-extraction cycles and the logical error rate per cycle, except that here the number of cycles is explicitly specified by \(n_{\mathrm{SE}}\).

A similar rule applies to faulty operations, except that they do not involve syndrome extraction:
{  \small
\begin{equation}
    \inferrule{\cS:= \cS^\Gamma_{q[i]}\\ F=(a,t,p,n_{\mathrm{anc}})\in \mathrm{OP}_F\in \cS\\ \epsilon:=\epsilon^{\Gamma}_{q[i]}\oplus\sem{p}_\cS\\T:=\sem{t}_\cS\\ N:=\sem{n_{\mathrm{anc}}}_\cS+\mathsf{AncSE}_\Gamma(\dom(\Gamma)\setminus\{q[i]\})}{\pare{F[q[i][j_1],\ldots,q[i][j_a]],\Gamma}\Downarrow\pare{\mathsf{Step}_\Gamma(\{q[i]\},\epsilon,T),T,N}}{},
\end{equation}
}

\noindent where the error rate \(p\) and time cost \(T\) are specified in \(\cS:=\cS^\Gamma_{q[i]}\).

The state preparation statement \(\treset(q[i])\) is treated analogously to a primitive logical operation, except that its error rate is \emph{refreshed} rather than accumulated, since the codeblock is reinitialized to a known state.
We have the following inference rule:
{  \small
\[
\inferrule{\cS:=\cS^\Gamma_{q[i]}\\ \mathrm{Prep}=(t,n_{\mathrm{SE}},n_{\mathrm{anc}})\in\cS\\
\mathrm{SE}=(n_{\mathrm{syn}},t_{\mathrm{SE}})\in\cS\\
\epsilon:=\sem{n_{\mathrm{SE}}}_\cS\sem{p_L}_\cS\\
T:=\sem{t}_\cS\\
N:=\sem{n_{\mathrm{anc}}}_\cS+\mathsf{AncSE}_\Gamma(\dom(\Gamma))}{\pare{\treset(q[i]),\Gamma}\Downarrow\pare{\mathsf{Step}_\Gamma(\{q[i]\},\epsilon,T),T,N}}{}.
\]
}

\noindent Intuitively, compared to the error context in Eq.~\eqref{eq:step}, the error rate of the active codeblock \(q[i]\) is reset to \(\epsilon:=\sem{n_{\mathrm{SE}}}_\cS\sem{p_L}_\cS\), rather than being accumulated on top of \(\epsilon^\Gamma_{q[i]}\).

For statement \(\tlogical\ q[m]:=\tnew\ \<{\tt QEC}\>;S;\trelease(q)\) and \(\cS\) specifying the error-correction scheme \(\<{\tt QEC}\>\), we have the following inference rule:
{  \small
\[
\inferrule{
    [[n,k,d]]\in \cS \\
    \Gamma_{\mathrm{ex}}:=\Gamma,q[0]:_0\cS,\ldots,q[m-1]:_0\cS\\
    \pare{S,\Gamma_{\mathrm{ex}}}\Downarrow\pare{\Gamma',t',n'}
}{\pare{\tlogical\ q[m]:=\tnew\ \<{\tt QEC}\>;S;\trelease(q),\Gamma}\Downarrow\pare{\Gamma'|_{\dom(\Gamma)},t',nm+n'}}{}.
\]
}



%
\noindent As illustrated in Fig.~\ref{fig:compilation-1}, the time cost is given by that of executing \(S\), while the space cost consists of the physical qubits allocated for the new codeblocks, namely \(nm\), together with the space cost \(n'\) incurred by executing \(S\).

For the error context, \(\Gamma\) is first extended to \(\Gamma_{\mathrm{ex}}\) by introducing the newly allocated codeblocks \(q[0],\ldots,q[m-1]\), each associated with the error-correction scheme \(\cS\) and initialized with error probability \(0\).
The statement \(S\) is then executed inductively under \(\Gamma_{\mathrm{ex}}\), yielding an intermediate error context \(\Gamma'\).
Finally, the resulting context is restricted to \(\dom(\Gamma)\) by discarding the newly allocated codeblocks, producing the final error context.

\subsection{Measurement-based Conditional Statements}
\begin{wrapfigure}{r}{0.34\textwidth}
    \centering
    \vspace{-2em}
    \includegraphics[scale=2]{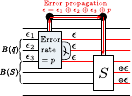}
    \caption{Error propagation in a measurement based \(\tif\) statement.}
    \label{fig:error-propagation}
\end{wrapfigure}
The measurement-based conditional statement is the most subtle case in resource estimation, since the measurement \(\cM\) may be either a specified primitive operation \(\cM=\cM(M)\) or a cross-code joint Pauli measurement \(\cM=\cM(P)\), in which errors may propagate across different codeblocks.
Moreover, the measurement outcome is available only after classical decoding, whose latency may cause additional error accumulation.
An incorrect measurement outcome may further propagate errors to the codeblocks involved in \(S\).

We first present a general framework for reasoning about error propagation in a measurement-based conditional statement.
As illustrated in Fig.~\ref{fig:error-propagation}, for a measurement with logical error rate \(p\), the error rate of the target codeblocks is conservatively updated to \(\bigoplus_i \epsilon_i \oplus p\), which captures the probability that all participating codeblocks are ideal prior to the measurement and that the measurement itself is error-free, accounting for error propagation within the measurement.

Furthermore, measurement errors may propagate to codeblocks outside the participating set \(\ActBlock(\bar{q})\), since the execution of \(S\) depends on the measurement outcome.
Therefore, the error rates of codeblocks in \(\ActBlock(S)\setminus \ActBlock(\bar{q})\) are additionally composed with an \(\epsilon\) to account for the possibility of an incorrect measurement outcome.
This also reflects that errors in \(\ActBlock(\bar{q})\) prior to the measurement may be propagated to codeblocks in \(\ActBlock(S)\setminus \ActBlock(\bar{q})\) and accumulated in their error rates.

\subsubsection{Specified primitive measurement.}
Suppose that the measurement is a specified primitive operation, which may be either logical or faulty.
If the measurement is faulty, then no decoding is required, and we obtain the following inference rule:
{  \small
\[
\inferrule{
    \cS:=\cS^\Gamma_{q[i]}\\
    M=(a,t,p,n_{\mathrm{anc}})\in \mathrm{OP}_F\in \cS\\
    \epsilon:=\epsilon^{\Gamma}_{q[i]}\oplus\sem{p}_\cS\\
    T:=\sem{t}_\cS\\
    N:=\sem{n_{\mathrm{anc}}}_{\cS}+\mathsf{AncSE}_\Gamma(\dom(\Gamma)\setminus\{q[i]\})\\
    (S_0,\mathsf{Step}_\Gamma(\{q[i]\},\epsilon,T))\Downarrow(\Gamma_1,t_1,n_1)
    \\
    (S,\Gamma_1)\Downarrow(\Gamma_2,t_2,n_2) \\
    \Gamma_{\mathrm{final}}:=\mathsf{Upd}_{\Gamma_2}\pare{\ActBlock(S)\setminus \{q[i]\},\epsilon}
}{
    \pare{
        \tif\ \cM(M)[q[i][j_1],\ldots,q[i][j_a]];
        S_0\ \tthen\ S,
        \Gamma
    }
    \Downarrow
    \pare{
        \Gamma_{\mathrm{final}},
        t_1+t_2+T,
        \max\left \{n_1,n_2,N\right \}
    }
}{}.
\]
}

\noindent Here, the time cost is simply the sum of the time required to perform the measurement and the time required to execute \(S_0\) and \(S\), as illustrated in Fig.~\ref{fig:compilation-5}.
The space cost is the maximum of the space costs of the measurement, \(S_0\), and \(S\), since these components are executed sequentially and may reuse ancillary qubits.

For error analysis, we first apply the faulty measurement operation to update the error context, obtaining \(\mathsf{Step}_\Gamma(\{q[i]\},\epsilon,T)\) as defined in Eq.~\eqref{eq:step}.
Statements \(S_0\) and \(S\) are then executed inductively, yielding \(\Gamma_1\) and \(\Gamma_2\), respectively.
Finally, since errors occurring during the measurement may affect the execution of \(S\), we conservatively update the final error context as
\(
\Gamma_{\mathrm{final}} := \mathsf{Upd}_{\Gamma_2}\pare{\ActBlock(S)\setminus \{q[i]\},\epsilon}.
\)
Here, \(\mathsf{Upd}_\Gamma(\cB,\epsilon)\) coincides with \(\Gamma\) except that, for each codeblock \(Q \in \cB\),
\[
\epsilon^{\mathsf{Upd}_{\Gamma}\pare{\cB,\epsilon}}_Q
=
\epsilon^{\Gamma}_Q \oplus \epsilon,
\] 
indicating that the codeblocks in \(\ActBlock(S)\setminus \{q[i]\}\) remain intact only if the measurement is error-free.

If the measurement is logical, then the decoding process is required to determine the measurement outcome, and we have the following inference rule:
{  \small
\[
\inferrule{
    \cS:=\cS^\Gamma_{q[i]}\\
    M=(a,t,n_{\mathrm{SE}},n_{\mathrm{anc},M})\in \mathrm{OP}_L\in \cS\\
    \mathrm{SE}=(n_{\mathrm{syn}},t_{\mathrm{SE}})\in\cS\\
    t_{\mathrm{dec}}\in\cS\\
    p_L\in\cS\\
    \epsilon:=\epsilon^\Gamma_{q[i]}\oplus\sem{n_{\mathrm{SE}}}_\cS\sem{p_L}_\cS\\
    T:=\sem{t}_\cS\\
    N:=\sem{n_{\mathrm{anc},M}}_\cS+\mathsf{AncSE}_\Gamma(\dom(\Gamma))\\
    (S_0,\mathsf{Step}_{\Gamma}\pare{\{q[i]\},\epsilon,T})\Downarrow (\Gamma_1,t_1,n_1)\\
    \pare{S,\mathsf{Wait}_{\Gamma_1}\pare{\sem{t_{\mathrm{dec}}}_\cS-t_1}^+}\Downarrow(\Gamma_2,t_2,n_2)\\
    \Gamma_{\mathrm{final}}:=\mathsf{Upd}_{\Gamma_2}\pare{\ActBlock(S)\setminus\{q[i]\},\epsilon}
}{\pare{
        \tif\ \cM(M)[q[i][j_1],\ldots,q[i][j_a]];
        S_0\ \tthen\ S,
        \Gamma
    }
    \Downarrow
    \pare{
        \Gamma_{\mathrm{final}},
        T+\max\left \{t_1,\sem{t_{\mathrm{dec}}}_{\cS}\right \}+t_2,
        \max\left \{n_1,n_2,N\right \}}
    }{},
\]
}

\noindent where \(x^+:=\max\{x,0\}\).
The time cost is the sum of the time required to perform the measurement, the time required to execute \(S\), and the maximum of the time required to execute \(S_0\) and the decoding time, since \(S_0\) and the decoding can be performed in parallel; and the space cost is the same as for faulty measurements with additional ancillary qubits for syndrome extraction.

For error analysis, we first update the error context by applying the logical measurement operation, obtaining \(\mathsf{Step}_\Gamma\pare{\{q[i]\},\epsilon,T}\).
We then execute \(S_0\) inductively and obtain \(\Gamma_1\).
Unlike the case of faulty measurement, however, the error context \(\Gamma_1\) is not used directly to execute \(S\), because the measurement outcome is available only after decoding is completed.
Therefore, we introduce an intermediate error context \(\mathsf{Wait}_{\Gamma_1}\pare{\sem{t_{\mathrm{dec}}}_\cS-t_1}^+\).
Finally, the error context is updated to \(\mathsf{Upd}_{\Gamma_{2}}\pare{\ActBlock(S)\setminus\{q[i]\},\epsilon}\) to account for the possibility that errors during the measurement may affect the subsequent execution of \(S\), as in the case of faulty measurement.

Intuitively, this construction accounts for the idle period between the completion of \(S_0\) and the completion of decoding, corresponding to the part labeled ``waiting for decoding'' in Fig.~\ref{fig:compilation-5}.
The length of this period is given by \(\pare{\sem{t_{\mathrm{dec}}}_{\cS}-t_1}^+\), since decoding may finish before \(S_0\) completes, in which case no idle period occurs and no additional errors are accumulated.

\subsubsection{Joint Pauli measurement.}
\label{sec:joint-pauli-re}
A joint Pauli measurement \(\cM(P)[\bar{q}]\) measures the eigenvalue of a multi-qubit Pauli operator \(P=P_1\otimes\cdots\otimes P_a\) on a list of distinct logical qubits
\[
    \bar{q}=(q_1[i_1][j_1],\ldots,q_a[i_a][j_a]).
\]
Unlike a primitive measurement specified inside a single error-correction
scheme, such a measurement may involve several codeblocks and these codeblocks may even be encoded by different schemes.
In our framework, \(\cM(P)[\bar{q}]\) is not treated as a specified operation.
Instead, we assume that it is implemented using the universal-adapter construction for quantum LDPC codes~\cite{1g44-jp62}, with its resource cost computed from the parameters of the participating codeblocks.

Briefly, for each codeblock of distance \(d_\ell\) with \(1\leq \ell\leq a\) that participates in the logical Pauli operator \(P\), one constructs an auxiliary graph whose edge qubits, together with its vertex and cycle checks, define the auxiliary code used for auxiliary-graph surgery.
The ports of these auxiliary graphs are then connected by adapter edges, thereby stitching together codeblocks associated with different error-correction codes into a single deformed codeblock of distance given by~\cite{1g44-jp62}:
\[
d_{\mathrm{joint}}=\min_{1\leq \ell\leq a} d_\ell.
\]
The joint Pauli measurement can then be performed as an intra-codeblock operation on this deformed codeblock.
In the following, we abstract away from the concrete implementation details and instead give a practically motivated estimate of the required resources.


The total number of ancillary qubits to implement auxiliary graph surgery for each codeblock, and the adapters between their auxiliary graphs, is estimated by
\begin{equation}\label{eq:anc-joint}
N_{\mathrm{joint}}=\mathsf{AncSE}_\Gamma\pare{\dom(\Gamma)}+\sum_{\ell =1}^a \sem{n_{\mathrm{aux}}^{(\ell)}}_{\cS^{(\ell)}}+(a-1)\cdot 3d_{\mathrm{joint}}.
\end{equation}
The first term in Eq.~\eqref{eq:anc-joint} accounts for the native syndrome qubits required by all codeblocks, regardless of whether they participate.
The second term corresponds to the ancillary qubits required by the auxiliary graph surgery associated with each individual codeblock, where \(n_{\mathrm{aux}}^{(\ell)}\) is specified in the corresponding error-correction scheme \(\cS^{(\ell)}\).

The third term in Eq.~\eqref{eq:anc-joint} captures the adapter edges needed to connect the auxiliary graphs of the participating code blocks.
For \(a\) participating code blocks, at most \(a-1\) adapters are required, as implied by Eq.~(20) in the proof of Theorem~3 of~\cite{1g44-jp62}.
Since the size of each adapter scales linearly with the joint distance \(d_{\mathrm{joint}}\), and following the practical implementations reported in Tables~III and~IV of~\cite{1g44-jp62}, we heuristically upper-bound the total adapter contribution by
\(
(a-1)\cdot 3d_{\mathrm{joint}}.
\)

The time cost admits a relatively clean estimate, as the joint Pauli measurement requires only \(O(d_{\mathrm{joint}})\) rounds of syndrome extraction~\cite{1g44-jp62}, analogous to a logical operation on a single codeblock of distance \(d_{\mathrm{joint}}\).
However, auxiliary graph surgery makes each individual round more involved, as the deformed code contains additional auxiliary and adapter checks.
Because the deformed code is still low density, this extra complexity increases the duration of each syndrome-extraction cycle only by a constant factor.
To account for both this per-cycle overhead and the constant factors hidden in the repeated-measurement protocol, we use a conservative heuristic estimate for the time cost of the joint Pauli measurement:
\[
T_{\mathrm{meas}}
=
d_{\mathrm{joint}}\cdot \max_{1\leq \ell\leq a} 2\sem{t_{\mathrm{SE}}^{(\ell)}}_{\cS^{(\ell)}}.
\]
Here, \(t_{\mathrm{SE}}^{(\ell)}\) denotes the time cost of one syndrome-extraction cycle in the \(\ell\)-th codeblock's error-correction scheme \(\cS^{(\ell)}\).
The factor \(2\) accounts for the increased complexity of each cycle caused by the auxiliary graph surgery, as the additional checks introduced by the auxiliary graph do not exceed the number of checks in the original codeblock.

The decoding latency is more opaque to estimate.
As discussed in~\cite{1g44-jp62}, a modular decoder, such as that in~\cite{cross2025improvedqldpcsurgerylogical}, is expected to decompose the decoding problem for the deformed code into two spatially separated components: an LDPC decoding problem on the original code and a matching problem on the auxiliary graph.
However, when measuring a large product of logical operators, further decomposition of the decoding problem may be necessary in practice in order to achieve fast decoding.
Thus, the decoding cost depends on the weight \(a\) of the measured Pauli operator \(P\), as well as on the decoding latencies of the participating codeblocks.
Under the assumption that a fast modular decoder is available, we heuristically estimate the decoding time by
\[
T_{\mathrm{dec}}
=
a\cdot \max_{1\leq \ell\leq a}
\sem{t_{\mathrm{dec}}^{(\ell)}}_{\cS^{(\ell)}}.
\]
This estimate reflects a conservative upper bound: the decoding problem for the deformed code is assumed to be decomposable into at most \(a\) subproblems, and each subproblem is assumed to be no harder than the most expensive decoding problem among the participating codeblocks.

Finally, for the error analysis, unlike previously introduced statements, a joint Pauli measurement may \emph{propagate} errors between codeblocks.
As a conservative estimate, we assume a unified error rate of all participating codeblocks after the joint measurement, given by the union bound over them. Specifically, the unified error rate is given by
\[
\epsilon=\bigoplus_{1\leq \ell\leq a}\epsilon^{(\ell)}\oplus \pare{d_{\mathrm{joint}}\cdot \max_{1\leq \ell\leq a}\sem{p_L^{(\ell)}}_{\cS^{(\ell)}}},
\]
where \(\epsilon^{(\ell)}\) denotes the error probability of the \(\ell\)-th codeblock in \(\Gamma\) before the joint measurement, and \(p_L^{(\ell)}\) denotes the logical error rate per syndrome-extraction cycle specified in the \(\ell\)-th codeblock's error-correction scheme \(\cS^{(\ell)}\).

In summary, the resource relation of the \(\tif\) statement guarded by joint Pauli measurements is characterized by the following inference rule:
{  \small
\[
\inferrule{
    \pare{S_0,\mathsf{Step}_\Gamma\pare{\{q_1[i_1],\ldots,q_a[i_a]\},\epsilon,T_{\mathrm{meas}}}}\Downarrow\pare{\Gamma_1,t_1,n_1}\\
    \pare{S,\mathsf{Wait}_{\Gamma_1}\pare{T_{\mathrm{dec}}-t_1}^+}\Downarrow\pare{\Gamma_2,t_2,n_2}\\
    \Gamma_{\mathrm{final}}:=\mathsf{Upd}_{\Gamma_2}\pare{\ActBlock(S)\setminus\{q_1[i_1],\ldots,q_a[i_a]\},\epsilon}
}{
    \pare{
        \tif\ \cM(P)[q_1[i_1][j_1],\ldots,q_a[i_a][j_a]];S_0\ \tthen\ S,\Gamma
    }
    \Downarrow
    \pare{
        \Gamma_{\mathrm{final}},
        T_{\mathrm{meas}}+\max\{t_1,T_{\mathrm{dec}}\}+t_2,
        \max\{N_{\mathrm{joint}},n_1,n_2\}
    }
}{},
\]
}

\noindent where \(\cS^{(\ell)}=\cS^\Gamma_{q_{\ell}[i_\ell]}\) for \(1\leq \ell\leq a\).
Here, the error rates of the participating codeblocks are unified to a common value \(\epsilon\) after the joint measurement, while the remaining parts are handled in the same manner as in the logical measurement case described above.

\subsection{Sequential, Parallel Composition and Subcircuit Calls}
The inference rule for sequential composition is given by
{  \small
\[
\inferrule{
\pare{S_1,\Gamma}\Downarrow\pare{\Gamma_1,t_1,n_1}
\quad
\pare{S_2,\Gamma_1}\Downarrow\pare{\Gamma_2,t_2,n_2}
}{
\pare{S_1;S_2,\Gamma}
\Downarrow
\pare{\Gamma_2,t_1+t_2,\max\{n_1,n_2\}}
}{},
\]
}

\noindent where the time costs are additive, while the space cost is given by the maximum of the two, since the statements are executed sequentially and ancillary qubits can be reused.

For parallel composition, the two statements act on disjoint sets of blocks. Suppose \(\dom(\Gamma)=D_1\uplus D_2\) is divided into two disjoint sets, \(\ActBlock(S_1)\subseteq D_1\), \(\ActBlock(S_2)\subseteq D_2\), and \(t_1\leq t_2\) without loss of generality:
{  \small
\[
\inferrule{
\dom(\Gamma)=D_1\uplus D_2,\quad \ActBlock(S_1)\subseteq D_1,\quad \ActBlock(S_2)\subseteq D_2
\\
\pare{S_1,\Gamma|_{D_1}}\Downarrow\pare{\Gamma_1,t_1,n_1}
\\
\pare{S_2,\Gamma|_{D_2}}\Downarrow\pare{\Gamma_2,t_2,n_2}
\\
t_1\leq t_2
}{
\pare{S_1\parallel S_2,\Gamma}
\Downarrow
\pare{\mathsf{Wait}_{\Gamma_1}(t_2-t_1)\uplus\Gamma_2,t_2,\max\{n_1,\mathsf{AncSE}_\Gamma(D_1)\}+n_2}
}{}.
\]
}

\noindent The shorter branch waits for the longer branch to finish, and its blocks accumulate idle-memory error during the difference \(t_2-t_1\).
The space costs add because the two subprograms execute simultaneously and their auxiliary qubits cannot be shared.

Finally, subcircuit calls are handled by syntactic substitution of actual parameters for formal parameters:
{  \small
\[
\inferrule{
\tdef\ \<{\tt name}\>\pare{\bar{P}',\bar{q}'}\{S\}\mbox{ is declared}
\\
\pare{S[\bar{P}/\bar{P}',\bar{q}/\bar{q}'],\Gamma}\Downarrow\pare{\Gamma',t,n}
}{
\pare{\tcall\ \<{\tt name}\>(\bar{P},\bar{q}),\Gamma}
\Downarrow
\pare{\Gamma',t,n}
}{}.
\]
}


\section{Prototype Implementation and Case Studies}\label{sec:case}
In this section, we present a prototype implementation of our framework and three case studies designed to answer the following research questions:
\begin{enumerate}
\item [\textbf{RQ1}:]  
Can our framework enable more effective resource utilization and provide finer-grained resource analyses that better capture practical execution costs?
\item [\textbf{RQ2}:] Can our framework approximate the resource estimates obtained by case-by-case analyses, despite abstracting away implementation-specific details?

\item [\textbf{RQ3}:] Can our framework produce resource estimates that are consistent with those of widely used general-purpose estimation tools?
\end{enumerate}
Sec.~\ref{sec:case-logicalT} investigates the implementation of a logical \(T\) gate, whose resource costs are well understood, to address \textbf{RQ1};
Sec.~\ref{sec:case-distill} studies the 15-to-1 magic-state distillation circuit to address \textbf{RQ2}; and
Sec.~\ref{sec:case-compare} compares our estimates against those produced by Microsoft's QREv3~\cite{microsoft_qdk_qrev3} to address \textbf{RQ3}.
Finally, Sec.~\ref{sec:case-eea} studies the EEA subroutine
for ECDLP as an indicative reference for a practically important,
very-large-scale problem of broad interest.



\subsection{Logical \(T\) Gate via Magic-State Injection: Trade-offs Between Space and Time Costs}\label{sec:case-logicalT}
For the motivating example introduced in Sec.~\ref{sec:motex}, our estimator reports that its source code in Fig.~\ref{fig:logicalT-code}, instantiated with distance \(d=27\), requires approximately \(2247\) syndrome-extraction cycles and \(3724\) ancillary physical qubits, with an output logical error rate of approximately \(5.23\%\).
The space overhead, on the order of several thousand physical qubits, is caused by nested joint Pauli measurements and code surgery, and is broadly consistent with typical resource estimates for logical \(T\)-gate implementations operating in the \(10^{-2}\) logical-error regime under current hardware assumptions.
By contrast, the execution time is substantially higher than expected for a straightforward implementation, which motivates a more detailed analysis of the resource bottlenecks.

By enabling the command-line flag {\tt ----trace} and inspecting the detailed trace produced by our estimator, we find that this estimate is reasonable for two reasons:
(1) the decoding latency of two-block joint Pauli measurements is substantial and often dominates the execution time, with each measurement typically requiring hundreds of syndrome-extraction cycles; and
(2) the logical \(T\) gate is implemented straightforwardly via magic-state injection, without optimization or accounting for amortized time cost when embedded in a larger circuit.

\begin{figure}[t]
    \centering
    \includegraphics[width=\textwidth]{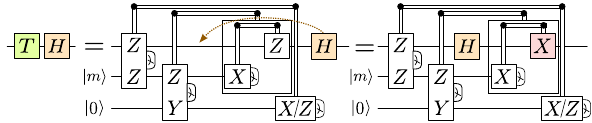}
    \caption{Auto-corrected implementation of a logical \(T\) gate using an additional ancilla, followed only by a Pauli correction. This form allows a subsequent Clifford gate to commute through the Pauli correction.}
    \label{fig:logicalT-auto}
\end{figure}

To address these limitations, we next show that, under appropriate optimizations and in an amortized sense, the time cost of a logical \(T\) gate can be significantly reduced.
An auto-corrected implementation is presented in Fig.~17(b) of~\cite{Litinski2019gameofsurfacecodes}, where the \(S\)-correction in Fig.~\ref{fig:logicalT} is eliminated at the cost of an additional ancilla.
As a result, only a Pauli correction is required, allowing subsequent Clifford gates to commute with the correction and thereby utilize the decoding latency more effectively.
For example, as illustrated in Fig.~\ref{fig:logicalT-auto}, a Hadamard gate \(H\) can be moved before the Pauli correction using \(HZ = XH\).

\begin{figure}[t]
    \centering
    \vspace{-1em}

\pgfplotstableread{
d   dec
23  0
25  0
27  255
29  470
31  1760
}\plottdecpercentile

\pgfplotstableread{
d   errA   timeA  spaceA  errB   timeB  spaceB
23   7.07  1150   2712    9.11  1058   3769
25   4.10  1250   3198    5.29  1150   4447
27   4.93  2544   3724    3.66  1752   5181
29   3.94  3713   4290    2.34  2274   5971
31   6.02  10257  4896    3.81  7350   6817
}\plotdataoddpercentile

\pgfplotstableread{
d   errA   timeA  spaceA  errB   timeB  spaceB
24   4.76  1226   2904    7.57  1132   3961
26   5.56  2269   3406    4.37  1224   4655
28   4.57  3478   3948    2.82  1826   5405
30   6.77  8947   4530    2.34  4135   6211
}\plotdataevenpercentile

\pgfplotstableread{
d   dec
23  200
25  280
27  390
29  520
31  680
}\plottdecmean

\pgfplotstableread{
d   errA   timeA  spaceA  errB   timeB  spaceB
23  13.83  2081   2712   10.74  1458   3769
25   9.38  2575   3198    6.55  1710   4447
27   6.37  3219   3724    3.99  2022   5181
29   4.23  3963   4290    2.40  2374   5971
31   2.78  4857   4896    1.61  3030   6817
}\plotdataoddmean

\pgfplotstableread{
d   errA   timeA  spaceA  errB   timeB  spaceB
24  10.94  2475   2904    8.45  1532   3961
26   7.45  3089   3406    5.03  1784   4655
28   4.97  3813   3948    3.00  2096   5405
30   3.27  4677   4530    1.76  2448   6211
}\plotdataevenmean

  \begin{tikzpicture}

    \pgfplotsset{
        oddA/.style={
            blue,
            solid,
            mark=*,
        },
        oddB/.style={
            red,
            solid,
            mark=square*,
        },
        evenA/.style={
            blue,
            mark=*,
            opacity=0.2,
            mark size=1.2pt,
        },
        evenB/.style={
            red,
            opacity=0.2,
            mark=square*,
            mark size=1.2pt,
        },
    }

    \begin{groupplot}[
        group style={
            group name=resourceplots,
            group size=4 by 1,
            horizontal sep=3.3em,
        },
        width=0.27\textwidth,
        height=0.26\textwidth,
        xlabel={Code distance $d$},
        xtick={23,25,27,29,31},
        xlabel style={yshift=0.5em},
        grid=both,
        tick label style={font=\scriptsize},
        label style={font=\scriptsize},
        title style={font=\scriptsize},
        legend style={
            font=\scriptsize,
            at={(1.3,-0.45)},
            anchor=north,
            legend columns=4,
        },
    ]

    \nextgroupplot[
        title={Mean decoding time},
        ylabel={\(\#\mathrm{SE}\) cycles},
        ylabel style={yshift=-0.5em},
        ymin=0,
        scaled y ticks=false,
        yticklabel={
            \pgfmathparse{\tick/100}
            \pgfmathprintnumber[fixed,precision=2]{\pgfmathresult}
        },
        extra description/.code={
            \node[font=\scriptsize, anchor=south west]
              at (axis description cs:-0.13,0.96) {$\times 10^2$};
        },
    ]
    \addplot+[mark=square*, red] table[x=d,y=dec] {\plottdecmean};
    \addplot+[mark=*, blue] table[x=d,y=dec] {\plottdecmean};

    \nextgroupplot[
        title={Error rate},
        ylabel={Error rate (\%)},
        ylabel style={yshift=-0.5em},
        ymin=0,
    ]
    \addplot+[oddA]  table[x=d,y=errA] {\plotdataoddmean};
    \addplot+[oddB]  table[x=d,y=errB] {\plotdataoddmean};
    \addplot+[evenA] table[x=d,y=errA] {\plotdataevenmean};
    \addplot+[evenB] table[x=d,y=errB] {\plotdataevenmean};

    \nextgroupplot[
        title={Space cost},
        ylabel={\#physical qubits},
        ylabel style={yshift=-0.5em},
        scaled y ticks=false,
        ytick distance=1000,
        yticklabel={
            \pgfmathparse{\tick/1000}
            \pgfmathprintnumber[fixed,precision=1]{\pgfmathresult}
        },
        extra description/.code={
            \node[font=\scriptsize, anchor=south west]
              at (axis description cs:-0.13,0.96) {$\times 10^3$};
        },
    ]
    \addplot+[oddA]  table[x=d,y=spaceA] {\plotdataoddmean};
    \addplot+[oddB]  table[x=d,y=spaceB] {\plotdataoddmean};
    \addplot+[evenA] table[x=d,y=spaceA] {\plotdataevenmean};
    \addplot+[evenB] table[x=d,y=spaceB] {\plotdataevenmean};

    \nextgroupplot[
        title={Time cost},
        ylabel={Total \(\#\mathrm{SE}\) cycles},
        ylabel style={yshift=-0.5em},
        ymin=0,
        ymax=11000,
        ytick={0,2500,5000,7500,10000},
        scaled y ticks=false,
        yticklabel={
            \pgfmathparse{\tick/1000}
            \pgfmathprintnumber[fixed,precision=1]{\pgfmathresult}
        },
        extra description/.code={
            \node[font=\scriptsize, anchor=south west]
              at (axis description cs:-0.13,0.96) {$\times 10^3$};
            \node[font=\scriptsize, anchor=south west]
              at (axis description cs:0.87,0.96) {$\times 10^2$};
        },
    ]
    \addplot+[oddA]  table[x=d,y=timeA] {\plotdataoddmean};
    \addplot+[oddB]  table[x=d,y=timeB] {\plotdataoddmean};
    \addplot+[evenA] table[x=d,y=timeA] {\plotdataevenmean};
    \addplot+[evenB] table[x=d,y=timeB] {\plotdataevenmean};

    \end{groupplot}

    \begin{axis}[
        at={(resourceplots c4r1.south west)},
        anchor=south west,
        width=0.27\textwidth,
        height=0.26\textwidth,
        xmin=23,
        xmax=31,
        ymin=0,
        ymax=11000,
        xtick=\empty,
        axis x line=none,
        axis y line*=right,
        ytick={0,2500,5000,7500,10000},
        scaled y ticks=false,
        yticklabel={
            \pgfmathparse{\tick/3700}
            \pgfmathprintnumber[fixed,precision=1]{\pgfmathresult}
        },
        ylabel={\(\#\mathrm{SE}\) cycles per gate},
        ylabel style={yshift=0.5em},
        tick label style={font=\scriptsize},
        label style={font=\scriptsize},
        grid=none,
    ]
    \end{axis}

    \end{tikzpicture}

    \begin{tikzpicture}

    \pgfplotsset{
        oddA/.style={
            blue,
            solid,
            mark=*,
        },
        oddB/.style={
            red,
            solid,
            mark=square*,
        },
        evenA/.style={
            blue,
            mark=*,
            opacity=0.2,
            mark size=1.2pt,
        },
        evenB/.style={
            red,
            opacity=0.2,
            mark=square*,
            mark size=1.2pt,
        },
    }

    \begin{groupplot}[
        group style={
            group name=resourceplots,
            group size=4 by 1,
            horizontal sep=3.3em,
        },
        width=0.27\textwidth,
        height=0.26\textwidth,
        xlabel={Code distance $d$},
        xtick={23,25,27,29,31},
        xlabel style={yshift=0.5em},
        grid=both,
        tick label style={font=\scriptsize},
        label style={font=\scriptsize},
        title style={font=\scriptsize},
        legend style={
            font=\scriptsize,
            at={(1.3,-0.45)},
            anchor=north,
            legend columns=4,
        },
    ]

    \nextgroupplot[
        title={$99.9^{\mathrm{th}}$ percentile decoding time},
        ylabel={\(\#\mathrm{SE}\) cycles},
        ylabel style={yshift=-0.5em},
        scaled y ticks=false,
        yticklabel={
            \pgfmathparse{\tick/100}
            \pgfmathprintnumber[fixed,precision=2]{\pgfmathresult}
        },
        extra description/.code={
            \node[font=\scriptsize, anchor=south west]
              at (axis description cs:-0.13,0.96) {$\times 10^2$};
        },
    ]
    \addplot+[mark=square*, red] table[x=d,y=dec] {\plottdecpercentile};
    \addplot+[mark=*, blue] table[x=d,y=dec] {\plottdecpercentile};

    \nextgroupplot[
        title={Error rate},
        ylabel={Error rate (\%)},
        ylabel style={yshift=-0.5em},
        ymin=0,
    ]
    \addplot+[oddA]  table[x=d,y=errA] {\plotdataoddpercentile};
    \addplot+[oddB]  table[x=d,y=errB] {\plotdataoddpercentile};
    \addplot+[evenA] table[x=d,y=errA] {\plotdataevenpercentile};
    \addplot+[evenB] table[x=d,y=errB] {\plotdataevenpercentile};
    \legend{\mbox{Naive\quad},  \mbox{Auto-corrected\quad}, \mbox{Naive hybrid\quad}, \mbox{Auto-corrected hybrid\quad}}

    \nextgroupplot[
        title={Space cost},
        ylabel={\#physical qubits},
        ylabel style={yshift=-0.5em},
        scaled y ticks=false,
        ytick distance=1000,
        yticklabel={
            \pgfmathparse{\tick/1000}
            \pgfmathprintnumber[fixed,precision=1]{\pgfmathresult}
        },
        extra description/.code={
            \node[font=\scriptsize, anchor=south west]
              at (axis description cs:-0.13,0.96) {$\times 10^3$};
        },
    ]
    \addplot+[oddA]  table[x=d,y=spaceA] {\plotdataoddpercentile};
    \addplot+[oddB]  table[x=d,y=spaceB] {\plotdataoddpercentile};
    \addplot+[evenA] table[x=d,y=spaceA] {\plotdataevenpercentile};
    \addplot+[evenB] table[x=d,y=spaceB] {\plotdataevenpercentile};

    \nextgroupplot[
        title={Time cost},
        ylabel={Total \(\#\mathrm{SE}\) cycles},
        ylabel style={yshift=-0.5em},
        ymin=0,
        ymax=11000,
        ytick={0,2500,5000,7500,10000},
        scaled y ticks=false,
        yticklabel={
            \pgfmathparse{\tick/1000}
            \pgfmathprintnumber[fixed,precision=1]{\pgfmathresult}
        },
        extra description/.code={
            \node[font=\scriptsize, anchor=south west]
              at (axis description cs:-0.13,0.96) {$\times 10^3$};
            \node[font=\scriptsize, anchor=south west]
              at (axis description cs:0.87,0.96) {$\times 10^2$};
        },
    ]
    \addplot+[oddA]  table[x=d,y=timeA] {\plotdataoddpercentile};
    \addplot+[oddB]  table[x=d,y=timeB] {\plotdataoddpercentile};
    \addplot+[evenA] table[x=d,y=timeA] {\plotdataevenpercentile};
    \addplot+[evenB] table[x=d,y=timeB] {\plotdataevenpercentile};

    \end{groupplot}

    \begin{axis}[
        at={(resourceplots c4r1.south west)},
        anchor=south west,
        width=0.27\textwidth,
        height=0.26\textwidth,
        xmin=23,
        xmax=31,
        ymin=0,
        ymax=11000,
        xtick=\empty,
        axis x line=none,
        axis y line*=right,
        ytick={0,2500,5000,7500,10000},
        scaled y ticks=false,
        yticklabel={
            \pgfmathparse{\tick/3700}
            \pgfmathprintnumber[fixed,precision=1]{\pgfmathresult}
        },
        ylabel={\(\#\mathrm{SE}\) cycles per gate},
        ylabel style={yshift=0.5em},
        tick label style={font=\scriptsize},
        label style={font=\scriptsize},
        grid=none,
    ]
    \end{axis}

    \end{tikzpicture}

    \caption{Resource estimation for a logical \(T\) gate followed by 36 logical
    Hadamard gates across different code distances, comparing the naive and
    auto-corrected implementations shown in Figs.~\ref{fig:logicalT}
    and~\ref{fig:logicalT-auto}, respectively.
    The two decoding-latency plots in the first column are adapted from Fig.~2(b) of~\cite{delfosse2023choosedecoderfaulttolerantquantum}.}
    \label{fig:resource-result}
  \end{figure}
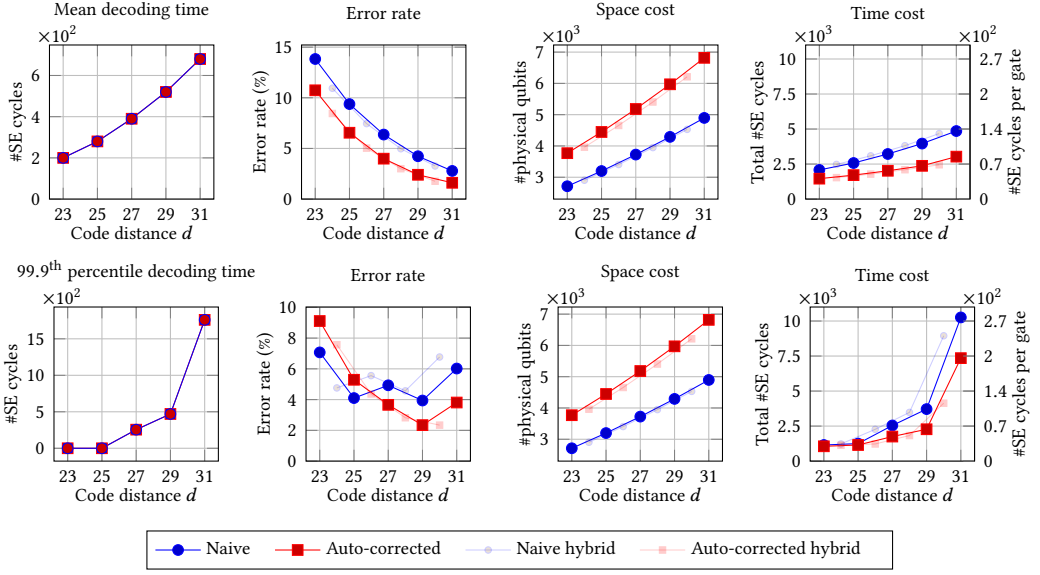

Thus, logical \(T\)-gate implementations involve an inherent trade-off among space cost, time cost, and error rate.
To investigate this trade-off, we consider an experiment in which a logical \(T\) gate is followed by 36 logical Hadamard gates, and compare the naive and optimized auto-corrected implementations across different code distances.
The results are shown in Fig.~\ref{fig:resource-result}, where the two rows correspond to mean and 99.9th-percentile estimates of the decoding time, respectively.

The more prominent data points at odd values of \(d\) correspond to the uniform setting where all codeblocks use the same code distance.
The more transparent points at even values of \(d\) represent a hybrid implementation, where data qubits and ancillary qubits use different code distances, namely distance \(d+1\) for data qubits and \(d-1\) for ancillas.

For mean decoding time, both time and space costs increase monotonically with code distance.
Meanwhile, the auto-corrected implementation achieves a lower error rate and reduced time cost at the expense of higher space cost, reflecting the expected resource trade-off.
However, under the 99.9th-percentile latency, whose variation is more pronounced, the error rate exhibits more intricate behavior: increasing code distance improves error-correction capability, but the accompanying increase in decoding time can also accumulate more memory errors.

This trade-off is particularly evident when the decoding time increases significantly, offsetting the gain in error-correction capability and yielding a higher error rate for the naive implementation.
In contrast, for the auto-corrected implementation, the decoding latency can be partially amortized, leading to an overall decreasing trend in the error rate as the code distance increases.
Nevertheless, when the latency surpasses the time required for the 36 logical Hadamard gates, the error rate rebounds.
These results answer \textbf{RQ1}: for a logical \(T\) gate whose resource costs are well understood and typically involve thousands of physical qubits and tens of syndrome-extraction cycles, our framework provides fine-grained resource analyses that capture realistic trade-offs among time, space, and error rates.

\begin{table}[t]
  \centering
  \caption{Mixed-distance strategies achieving optimal error rates under different space budgets.}
  \label{tab:optimal}
  \begin{tabular}{c c c c}
  \toprule
  \multicolumn{1}{c}{Space budget}
  & \multicolumn{1}{c}{Optimal error rate}
  & \multicolumn{1}{c}{Strategy}
  & \multicolumn{1}{c}{Time cost} \\
  \midrule

  $<5000$ qubits & $6.6387\%$
  & $(d_q = 25,\ d_{\mathrm{anc}[0]} = 25,\ d_{\mathrm{anc}[1]} = 23)$
  & $1620$ cycles \\

  $<5500$ qubits & $5.0343\%$
  & $(d_q = 27,\ d_{\mathrm{anc}[0]} = 25,\ d_{\mathrm{anc}[1]} = 25)$
  & $1784$ cycles \\

  $<6000$ qubits & $3.3453\%$
  & $(d_q = 29,\ d_{\mathrm{anc}[0]} = 27,\ d_{\mathrm{anc}[1]} = 23)$
  & $1886$ cycles \\

  $<6500$ qubits & $2.5067\%$
  & $(d_q = 29,\ d_{\mathrm{anc}[0]} = 29,\ d_{\mathrm{anc}[1]} = 25)$
  & $2114$ cycles \\

  $<7000$ qubits & $1.8129\%$
  & $(d_q = 31,\ d_{\mathrm{anc}[0]} = 29,\ d_{\mathrm{anc}[1]} = 27)$
  & $2308$ cycles \\

  $<7500$ qubits & $1.6618\%$
  & $(d_q = 31,\ d_{\mathrm{anc}[0]} = 31,\ d_{\mathrm{anc}[1]} = 27)$
  & $2720$ cycles \\

  \bottomrule
  \end{tabular}
\end{table}
By exploiting the flexibility of our language, a more nuanced trade-off between space cost and error rate can be achieved by allowing data and ancillary qubits with different code distances to coexist within the same circuit, as illustrated by the transparent data points at even values of \(d\) in Fig.~\ref{fig:resource-result}.
To further demonstrate the advantage of such hybrid implementations, we evaluate the optimal error rate under different space budgets for the mean-decoding-time case with the auto-corrected implementation, as summarized in Table~\ref{tab:optimal}.
The results indicate that the error-correction requirements of the data qubit {\tt q} and the ancillary qubits {\tt anc[0]} and {\tt anc[1]} are inherently asymmetric.
Under a constrained space budget, a lower error rate is achieved by prioritizing stronger protection for {\tt q}, while assigning a smaller code distance to {\tt anc[1]}.

\subsection{The 15-to-1 Magic State Distillation Protocol}\label{sec:case-distill}
\begin{figure}[t]
    \begin{minipage}{0.5\textwidth}
        \includegraphics[scale=1.2]{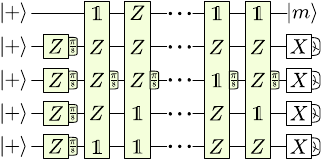}
        \caption{The 15-to-1 magic state distillation protocol, where a lighter-colored box indicates a faulty gate, in contrast to the logical \(T\) gate in the previous subsection.}
        \label{fig:distill-15to1}
    \end{minipage}
    \hspace{1em}
    \begin{minipage}{0.45\textwidth}
        \centering
        \includegraphics[scale=1.2]{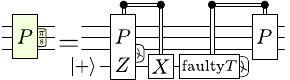}\\[0.5em]
        \includegraphics[scale=1.9]{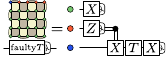}
        \caption{Implementation of a faulty \(T\) gate using an ancilla and a primitive faulty \(T\) measurement.}
        \label{fig:faultyT}
    \end{minipage}
\end{figure}

In this subsection, we study the preparation of the magic state \(|m\>\coloneqq T|+\>\) using the 15-to-1 distillation protocol presented in~\cite{Litinski2019magicstate}, which applies 15 faulty rotation gates to produce a single high-fidelity magic state as illustrated in Fig.~\ref{fig:distill-15to1}.
Each faulty rotation gate is implemented using an ancilla together with a primitive faulty \(T\)-measurement, as shown in Fig.~\ref{fig:faultyT}.

A protocol-specific resource analysis in the original paper~\cite{Litinski2019magicstate} shows that one execution of the circuit in Fig.~\ref{fig:distill-15to1} requires \(2\cdot (d_X+4d_Z)\cdot 3d_X + 4d_m\) physical qubits and approximately \(6d_m\) code cycles (see Fig.~11 of~\cite{Litinski2019magicstate}).
This estimate accounts for several implementation details and optimizations, including the
two-dimensional surface-code lattice-surgery layout, the use of separate
distance parameters \(d_X,d_Z,d_m\) to exploit the asymmetry among different
error mechanisms, and parallel execution together with pipelined reuse of
ancillary regions to reduce the overall space--time cost.

\begin{figure}[t]
        \begin{lstlisting}[style=ftqp]
def rotation3(P: Pauli, q1: Codeblock, q2: Codeblock, q3: Codeblock) {
  logical anc := new surface_code; if M(PZ)[q1, q2, q3, anc]; then X[anc]; if M(faultyT)[anc]; then P[q1, q2, q3]; release(anc);
}
def rotation1(P: Pauli, q: Codeblock) { ... }
def rotation5(P: Pauli, q1: Codeblock, q2: Codeblock, q3: Codeblock, q4: Codeblock, q5: Codeblock) { ... }

def distill_circuit(q: Codeblock) {
  logical anc[4] := new surface_code;
  prepare(anc[0]) || prepare(anc[1]) || prepare(anc[2]) || prepare(anc[3]);
  H[q] || H[anc[0]] || H[anc[1]] || H[anc[2]] || H[anc[3]];
  (call rotation1(Z, anc[0])) || (call rotation1(Z, anc[1])) || (call rotation1(Z, anc[2])) || (call rotation1(Z, anc[3]));
  call rotation3(ZZZ, anc[0], anc[1], anc[2]);
  ...... 
  call rotation3(ZZZ, anc[0], anc[1], anc[3]);
  (if M(X)[anc[0]]; then skip) || (if M(X)[anc[1]]; then skip) || (if M(X)[anc[2]]; then skip) || (if M(X)[anc[3]]; then skip);
  release(anc);
}
        \end{lstlisting}
    \caption{Source code for the 15-to-1 magic state distillation circuit.}
    \label{fig:distill-code}
\end{figure}
However, these protocol-specific optimizations can hardly be modeled in our general-purpose framework without hand-tuned calibration. 
To demonstrate the effectiveness of our framework, we do not deliberately tune the specifications to reproduce the reported results; instead, we use a more straightforward, though less optimized, implementation of the same circuit based on the source code shown in Fig.~\ref{fig:distill-code}, where the implementations of the individual rotation gates and intermediate operations are omitted for brevity.
To disable the optimization based on asymmetric error mechanisms, we set \(d=d_X=d_Z=d_m\) and perform resource estimation under the corresponding surface-code specifications for different values of \(d\), comparing the resulting estimates with those reported in~\cite{Litinski2019magicstate}.

The results are summarized in Fig.~\ref{fig:distill-compare}.
They show that our framework provides a close approximation of the space cost reported in~\cite{Litinski2019magicstate}, as the space cost is largely dominated by the number of logical qubits and their code parameters, with only minor discrepancies arising from layout constraints and adapters introduced by our generic compilation strategy.
The estimated time cost is higher in absolute terms, but follows the same asymptotic scaling as the estimates in~\cite{Litinski2019magicstate}; specifically, our framework yields a time cost of approximately \(52d+17\) code cycles, compared with \(6d\) code cycles in the protocol-specific construction, resulting in a constant-factor overhead of about \(8.7\times\) for large \(d\).
This overhead is mainly due to our unoptimized implementation, which does not exploit the pipelined execution and surgery reuse used in the protocol-specific construction.

Taken together, these results answer \textbf{RQ2}: our framework captures the dominant sources of resource consumption and produces estimates that are broadly aligned with case-by-case analyses, with discrepancies limited to constant factors, thereby demonstrating its reliability and practical applicability for resource estimation in large-scale fault-tolerant quantum computing.


\subsection{Calibration Against Microsoft's Quantum Resource Estimator}\label{sec:case-compare}
\pgfplotsset{
  compat=1.18,
  compareLine/.style={no marks, solid, opacity=0.75, line width=1.1pt},
  msPoint/.style={only marks, mark=o, mark size=2.8pt, line width=0.9pt, mark options={fill=white, line width=0.9pt}},
  ourPoint/.style={only marks, mark=x, mark size=2.8pt, line width=0.9pt, mark options={line width=0.9pt}},
}
\definecolor{qubitsHundred}{RGB}{12,74,120}
\definecolor{qubitsTwoHundred}{RGB}{156,76,0}
\definecolor{qubitsThreeHundred}{RGB}{20,96,42}
\definecolor{qubitsFourHundred}{RGB}{104,50,130}
\newcommand{\legendLine}[1]{\textcolor{#1}{\rule[0.35ex]{1.4em}{1.1pt}}}
\newcommand{\legendCircle}{\raisebox{-0.08ex}{\scriptsize$\circ$}}
\newcommand{\legendCross}{\raisebox{-0.08ex}{\scriptsize$\times$}}


\begin{figure}[t]
  \centering
  \vspace{-1em}

\begin{minipage}{0.40\textwidth}

\centering
    \begin{tikzpicture}
    \begin{groupplot}[
        group style={
            group size=2 by 1,
            horizontal sep=0.9cm,
        },
        width=0.25*396pt,
        height=0.25*396pt,
        x=0.3cm,
        enlarge x limits={abs=0.15cm},
        ybar,
        symbolic x coords={d1,d2,d3,d4,d5,d6},
        xtick=data,
        xticklabels={
            {$11$},
            {$21$},
            {$31$},
            {$41$},
            {$51$},
            {$61$}
        },
        x tick label style={font=\tiny, yshift=0.5em},
        xtick style={draw=none},
        xlabel={Code distance},
        ymin=0,
        grid=both,
        grid style={dashed,gray!30},
        y tick label style={font=\tiny},
        label style={font=\tiny},
        legend style={
            at={(-0.3,-0.7)},
            anchor=south,
            legend columns=2,
            draw=none,
            font=\scriptsize,
            fill=white,
            fill opacity=0.8,
            text opacity=1
        },
        title style={font=\scriptsize},
        xlabel style={yshift=0.5em},
        ylabel style={yshift=-0.5em},
    ]

    \nextgroupplot[
        title={Space cost},
        ylabel={\# physical qubits},
        ymode=log,
        log basis y=10,
        ymin = 100,
        y tick label style={
            font=\tiny,
            xshift=0.3em
        }
    ]

    \addplot+[
        fill=blue!50,
        draw=blue!80,
        /pgf/bar width=3.5pt,
        bar shift=-1pt
    ] coordinates {
        (d1,4237)
        (d2,15237)
        (d3,33037)
        (d4,57637)
        (d5,89037)
        (d6,127237)
    };

    \addplot+[
        fill=orange!55,
        draw=orange!90,
        /pgf/bar width=3.5pt,
        bar shift=1pt
    ] coordinates {
        (d1,3674)
        (d2,13314)
        (d3,28954)
        (d4,50594)
        (d5,78234)
        (d6,111874)
    };

    \nextgroupplot[
        title={Time cost},
        ylabel={\(\#\mathrm{SE}\) cycles},
        ymode=log,
        log basis y=10,
        ymin = 1,
        ymax = 4000,
        y tick label style={
            font=\tiny,
            xshift=0.3em
        }
    ]
    \addplot+[
        fill=blue!50,
        draw=blue!80,
        /pgf/bar width=3.5pt,
        bar shift=-1pt
    ] coordinates {
        (d1,589)
        (d2,1109)
        (d3,1629)
        (d4,2149)
        (d5,2669)
        (d6,3189)
    };
    \addplot+[
        fill=orange!55,
        draw=orange!90,
        /pgf/bar width=3.5pt,
        bar shift=1pt
    ] coordinates {
        (d1,66)
        (d2,126)
        (d3,186)
        (d4,246)
        (d5,306)
        (d6,366)
    };
    \legend{\mbox{Our framework\quad}, Reported in~\cite{Litinski2019magicstate}}
    \end{groupplot}
    \end{tikzpicture}
  \caption{Comparison of space and time costs at different code distances for the program in Fig.~\ref{fig:distill-code}.}
  \label{fig:distill-compare}
\end{minipage}
\hspace{1em}
\begin{minipage}{0.55\textwidth}
  \centering
\begin{tikzpicture}
\begin{groupplot}[
  group style={
    group name=resourceplots,
    group size=3 by 1,
    horizontal sep=2.0em,
  },
  width=0.25*396pt,
  height=0.25*396pt,
  xlabel={Time cost (ms)},
  ylabel={Space cost (qubits)},
  xlabel style={yshift=0.5em},
  ylabel style={yshift=-0.5em},
  grid=both,
  scaled x ticks=false,
  scaled y ticks=true,
  tick label style={font=\tiny},
  label style={font=\tiny},
  title style={font=\scriptsize},
]
\nextgroupplot[
  title={QFT},
]
\addplot+[compareLine, color=qubitsHundred, forget plot] coordinates {
  (18.711,212330)
  (80.0982,212769)
};
\addplot+[msPoint, color=qubitsHundred, forget plot] coordinates {
  (18.711,212330)
};
\addplot+[ourPoint, color=qubitsHundred, forget plot] coordinates {
  (80.0982,212769)
};
\addplot+[compareLine, color=qubitsTwoHundred, forget plot] coordinates {
  (37.611,398221)
  (160.738,399541)
};
\addplot+[msPoint, color=qubitsTwoHundred, forget plot] coordinates {
  (37.611,398221)
};
\addplot+[ourPoint, color=qubitsTwoHundred, forget plot] coordinates {
  (160.738,399541)
};
\addplot+[compareLine, color=qubitsThreeHundred, forget plot] coordinates {
  (56.511,582350)
  (241.378,583670)
};
\addplot+[msPoint, color=qubitsThreeHundred, forget plot] coordinates {
  (56.511,582350)
};
\addplot+[ourPoint, color=qubitsThreeHundred, forget plot] coordinates {
  (241.378,583670)
};
\addplot+[compareLine, color=qubitsFourHundred, forget plot] coordinates {
  (75.411,765598)
  (322.018,766478)
};
\addplot+[msPoint, color=qubitsFourHundred, forget plot] coordinates {
  (75.411,765598)
};
\addplot+[ourPoint, color=qubitsFourHundred, forget plot] coordinates {
  (322.018,766478)
};
\nextgroupplot[
  title={BV},
]
\addplot+[compareLine, color=qubitsHundred, forget plot] coordinates {
  (0,202630)
  (8.82,203132)
};
\addplot+[msPoint, color=qubitsHundred, forget plot] coordinates {
  (0,202630)
};
\addplot+[ourPoint, color=qubitsHundred, forget plot] coordinates {
  (8.82,203132)
};
\addplot+[compareLine, color=qubitsTwoHundred, forget plot] coordinates {
  (0,388521)
  (17.64,389904)
};
\addplot+[msPoint, color=qubitsTwoHundred, forget plot] coordinates {
  (0,388521)
};
\addplot+[ourPoint, color=qubitsTwoHundred, forget plot] coordinates {
  (17.64,389904)
};
\addplot+[compareLine, color=qubitsThreeHundred, forget plot] coordinates {
  (0,572650)
  (26.46,574033)
};
\addplot+[msPoint, color=qubitsThreeHundred, forget plot] coordinates {
  (0,572650)
};
\addplot+[ourPoint, color=qubitsThreeHundred, forget plot] coordinates {
  (26.46,574033)
};
\addplot+[compareLine, color=qubitsFourHundred, forget plot] coordinates {
  (0,755898)
  (35.28,756841)
};
\addplot+[msPoint, color=qubitsFourHundred, forget plot] coordinates {
  (0,755898)
};
\addplot+[ourPoint, color=qubitsFourHundred, forget plot] coordinates {
  (35.28,756841)
};
\nextgroupplot[
  title={Ripple-Carry Adder},
]
\addplot+[compareLine, color=qubitsHundred, forget plot] coordinates {
  (51.45,212330)
  (189.924,212769)
};
\addplot+[msPoint, color=qubitsHundred, forget plot] coordinates {
  (51.45,212330)
};
\addplot+[ourPoint, color=qubitsHundred, forget plot] coordinates {
  (189.924,212769)
};
\addplot+[compareLine, color=qubitsTwoHundred, forget plot] coordinates {
  (103.95,398221)
  (383.544,399541)
};
\addplot+[msPoint, color=qubitsTwoHundred, forget plot] coordinates {
  (103.95,398221)
};
\addplot+[ourPoint, color=qubitsTwoHundred, forget plot] coordinates {
  (383.544,399541)
};
\addplot+[compareLine, color=qubitsThreeHundred, forget plot] coordinates {
  (156.45,582350)
  (577.164,583670)
};
\addplot+[msPoint, color=qubitsThreeHundred, forget plot] coordinates {
  (156.45,582350)
};
\addplot+[ourPoint, color=qubitsThreeHundred, forget plot] coordinates {
  (577.164,583670)
};
\addplot+[compareLine, color=qubitsFourHundred, forget plot] coordinates {
  (208.95,765598)
  (770.784,766478)
};
\addplot+[msPoint, color=qubitsFourHundred, forget plot] coordinates {
  (208.95,765598)
};
\addplot+[ourPoint, color=qubitsFourHundred, forget plot] coordinates {
  (770.784,766478)
};

\end{groupplot}

\node[anchor=north, yshift=-2em, align=center, font=\scriptsize] at (resourceplots c2r1.south) {%
  $
  \begin{array}{c}
    \text{\legendLine{qubitsHundred}~100 qubits}\qquad
    \text{\legendLine{qubitsTwoHundred}~200 qubits}\qquad
    \text{\legendLine{qubitsThreeHundred}~300 qubits}\qquad
    \text{\legendLine{qubitsFourHundred}~400 qubits}\qquad\\
    \text{\legendCircle~Microsoft QREv3~\cite{microsoft_qdk_qrev3}}\qquad
    \text{\legendCross~Our Framework}
  \end{array}
  $
};
\end{tikzpicture}
\vspace{-1.8em}
  \caption{Comparison of space and time costs between Microsoft's Quantum Resource Estimator (Azure QREv3) and our framework for three practical circuits.}
  \label{fig:compare-ms-resource}

\end{minipage}

\end{figure}

To further validate the practical accuracy of our framework, we compare its resource-estimation results with those produced by Microsoft Quantum Resource Estimator~\cite{microsoft_resource_estimator} on realistic large-scale quantum circuits.
Specifically, we adopt the parameter settings from the documentation of its latest version available at the time of our experiments (QREv3) in~\cite{microsoft_qdk_qrev3}. The only modification is that we fix the code distance to \(d=21\), while keeping all other parameters unchanged, as shown below:
\begin{lstlisting}[language=Python, style=output]
app = OpenQASMApplication(source)
arch = GateBased(error_rate=1e-4, gate_time=100, measurement_time=500)
isa_query = SurfaceCode.q(distance=21) * RoundBasedFactory.q()
estimates = estimate(app, arch, isa_query, max_error=1e-4, name=qasm_path.name)
\end{lstlisting}

We then calibrate the specifications and configurations in our framework to match those of QREv3, including the error model, and treat the logical \(T\) gate as a primitive operation for consistency.
Finally, we perform the same estimation tasks on 100-, 200-, 300-, and 400-qubit instances of the quantum Fourier transform (QFT), the Bernstein--Vazirani (BV) algorithm, and Ripple-Carry Adder circuits. In the latter, Toffoli gates are decomposed into Clifford and \(T\) gates. 

The experimental results are presented in Fig.~\ref{fig:compare-ms-resource}. 
Overall, the space-cost estimates produced by our framework are almost perfectly aligned with those of QREv3, with the minor discrepancy attributable to the different treatments of logical CNOT gates: QREv3 models them as primitive operations, whereas our framework implements them via joint Pauli measurements and code surgery, thereby introducing additional ancillary qubits and a corresponding increase in space overhead. 
By contrast, the discrepancy in time cost is more pronounced, since QREv3 primarily accounts for the latency of logical \(T\) gates and assumes CNOT gates incur no time overhead, whereas our framework additionally incorporates the latency associated with joint Pauli measurements and classical decoding.
The comparison answers \textbf{RQ3}: across a variety of large-scale quantum circuits, our framework produces resource estimates that closely align with those produced by QREv3, a widely used general-purpose estimation tool, thereby supporting the practical accuracy of our conservative resource-estimation approach.

\subsection{Practice at Very Large Scale: the Extended Euclidean Algorithm}\label{sec:case-eea}
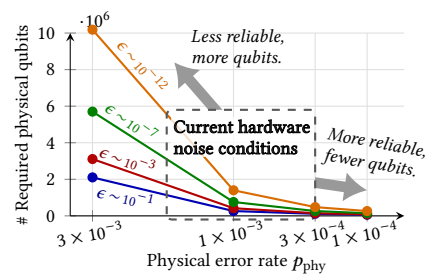
\begin{wrapfigure}{r}{0.4\textwidth}
    \vspace{-1em}
    \begin{tikzpicture}
\begin{axis}[
    width=6cm,
    height=4cm,
    clip=false,
    xlabel={Physical error rate \(p_{\mathrm{phy}}\)},
    ylabel={\# Required physical qubits},
    xmin=0,
    xmax=4.5,
    ymin=0,
    ymax=10000000,
    scaled y ticks=base 10:-6,
    xticklabels={
        {$3\times 10^{-3}$},
        {$1\times 10^{-3}$},
        {$3\times 10^{-4}$}, 
        {$1\times 10^{-4}$}
    },
    xtick={0.3,2.2,3.3,4.0},
    xticklabel style={font=\scriptsize, rotate=10, yshift=0.5em},
    xlabel style={font=\scriptsize, yshift=0.5em},
    ytick={0,2000000,4000000,6000000,8000000,10000000},
    yticklabel style={font=\scriptsize},
    ylabel style={font=\scriptsize, yshift=-0.5em},
    grid=both,
    tick align=outside,
    major grid style={gray!25},
    axis lines=left,
    legend columns=2,
    legend style={
        font=\scriptsize,
        at={(0.67,1.1)},
        anchor=north
    },
]

\addplot[
    thick,
    mark=*,
    mark size=1.5pt,
    color=blue!70!black,
    mark options={fill=blue!70!black}
]
coordinates {
    (0.3, {249*2*65^2})
    (2.2, {249*2*23^2})
    (3.3, {249*2*15^2})
    (4.0, {249*2*11^2})
};

\addplot[
    thick,
    mark=*,
    mark size=1.5pt,
    color=red!70!black,
    mark options={fill=red!70!black}
]
coordinates {
    (0.3, {249*2*79^2})
    (2.2, {249*2*29^2})
    (3.3, {249*2*17^2})
    (4.0, {249*2*13^2})
};

\addplot[
    thick,
    mark=*,
    mark size=1.5pt,
    color=green!50!black,
    mark options={fill=green!50!black}
]
coordinates {
    (0.3, {249*2*107^2})
    (2.2, {249*2*39^2})
    (3.3, {249*2*23^2})
    (4.0, {249*2*17^2})
};

\addplot[
    thick,
    mark=*,
    mark size=1.5pt,
    color=orange!85!black,
    mark options={fill=orange!85!black}
]
coordinates {
    (0.3, {249*2*143^2})
    (2.2, {249*2*53^2})
    (3.3, {249*2*31^2})
    (4.0, {249*2*23^2})
};




\node[align=center, anchor=south,text=orange!80!black, rotate=-50]
    at (axis cs:0.8, 7600000)
    {\scriptsize\(\epsilon\!\sim\!{\scriptscriptstyle 10}^{-12}\)};

\node[align=center, anchor=south,text=green!50!black, rotate=-35]
    at (axis cs:0.7,4400000)
    {\scriptsize\(\epsilon\!\sim\!{\scriptscriptstyle 10}^{-7}\)};

\node[align=center, anchor=south,text=red!50!black, rotate=-20]
    at (axis cs:0.7,2200000)
    {\scriptsize\(\epsilon\!\sim\!{\scriptscriptstyle 10}^{-3}\)};

\node[align=center, anchor=south,text=blue!50!black, rotate=-13]
    at (axis cs:0.7,160000)
    {\scriptsize\(\epsilon\!\sim\!{\scriptscriptstyle 10}^{-1}\)};
\def\xleft{1.3}
\def\xright{3.3}
\def\ybottom{-200000}
\def\ytop{5800000}
\begin{scope}
    \clip
        (current axis.south west)
            -- (current axis.south east)
            -- (current axis.north east)
            -- (current axis.north west)
            -- cycle
        (axis cs:\xleft,\ybottom)
            -- (axis cs:\xleft,\ytop)
            -- (axis cs:\xright,\ytop)
            -- (axis cs:\xright,\ybottom)
            -- cycle;
    \draw[
        -{Latex[length=3mm,width=4mm]},
        line width=4pt,
        draw=gray!80!white,
    ] (axis cs:3.0, 2.0e6) -- (axis cs:4.0,1.4e6);
    \draw[
        -{Latex[length=3mm,width=4mm]},
        line width=4pt,
        draw=gray!80!white,
    ] (axis cs:2.0, 5.6e6) -- (axis cs:1.4, 8.4e6);
\end{scope}
\draw[
    dashed,
    rounded corners=2pt,
    thick,
    color=gray!70!black,
]
(axis cs:1.3,-200000) rectangle (axis cs:3.3,5800000);
\node[
    anchor=south west,
    font=\scriptsize,
    align=left,
    fill=white,
    fill opacity=0.8,
    inner sep=1pt,
] at (axis cs:1.35, 3.2e6) {\pdfliteral direct {2 Tr 0.25 w}Current hardware\\ noise conditions\pdfliteral direct {0 Tr}};
\node[
    anchor=south west,
    font=\scriptsize\itshape,
    align=left,
    rotate=-10,
] at (axis cs:3.25, 2.4e6) {More reliable,\\fewer qubits.};
\node[
    anchor=south west,
    font=\scriptsize\itshape,
    align=left,
] at (axis cs:1.5, 7.6e6) {Less reliable,\\more qubits.};
\end{axis}
\end{tikzpicture}
    \vspace{-2em}
    \caption{Physical qubit requirements for \(\sim 10^9\)-cycle programs across hardware noise conditions and error budgets \(\epsilon\).}
    \label{fig:current}
\end{wrapfigure}
We additionally present a component-level study of a publicly
available implementation of the extended Euclidean algorithm (EEA) for
modular inversion~\cite{eea}, with the detailed setup deferred to the appendix. 
Although the target paper~\cite{eea} studies Shor's algorithm for elliptic-curve discrete logarithm problem (ECDLP), its released repository~\cite{eea-repo} provides only the EEA-based reversible modular-inversion subroutine, which the paper identifies as the dominant contributor to the logical-space cost of affine elliptic-curve point addition.

This case is intended to provide an indicative reference for a practically
important large-scale problem of broad interest, rather than a rigorous
benchmark, because the approximately two-million-line source
program is mechanically translated from the implementation
in~\cite{eea-repo}, and our analysis estimates only its resource
requirements without verifying the correctness of the original
implementation.
Running our estimator on this instance takes \(\sim 13\) minutes
and yields an estimate of approximately \(2.12\times 10^6\) physical qubits and
\(3.5\times 10^{9}\) cycles for the 64-bit EEA circuit, which uses
7-bit length registers and contains 371 EEA iterations.
These statistics are admittedly contrived, especially for the time cost, because the implementation in~\cite{eea-repo} prioritizes space efficiency and constructs with no gate parallelism, resulting in a relatively large cycle count.

To illustrate more intuitively how hardware quality affects the resource cost of such large-scale programs, Fig.~\ref{fig:current} reports the required number of physical qubits under varying hardware noise rates and error budgets. 
The estimates are computed as
\[
\#\mathrm{Required\ physical\ qubits}
=
\min\left\{
249\cdot 2d^2
\;:\;
10^9 \cdot 0.1
\left({p_{\mathrm{phy}}}/{0.57\%}\right)^{(d+1)/2}
< \epsilon
\right\},
\]
where the EEA circuit takes \(249\) logical qubits and \(\sim 10^9\) cycles.
Current state-of-the-art experimental hardware operates in the \(10^{-3}\)--\(10^{-4}\) physical-error-rate regime~\cite{google-nature,helios}, with only around one hundred physical qubits available.
The results indicate that each order-of-magnitude reduction in the physical error rate leads to an exponential decrease in the required number of physical qubits, highlighting a promising path from NISQ era toward large-scale FTQC.

\section{Discussion: Impact and Limitations}\label{sec:discussion}
Compared with existing methods, which typically estimate logical resources in an error-free setting before translating them into physical resources under error-correction parameters, our framework provides a more integrated analysis of time, space, and error rates.
It also allows programmers to assign heterogeneous error-correction schemes to different program components, enabling more efficient resource utilization and providing a basis for exploring resource-allocation strategies.

However, our framework currently focuses on program-level analysis and relies on an abstract execution model.
It does not yet capture lower-level compilation and scheduling constraints, such as layout and connectivity.
As a result, the estimates may be conservative, while higher precision may require programmers to carefully structure programs and specify error-correction schemes and hardware parameters.
\section{Related Work}\label{sec:rw}

\paragraph{Fault-tolerant quantum computing.}
Since the introduction of fault-tolerant quantum computing~\cite{FTQC} and the threshold theorem~\cite{FTQC-constant}, extensive research has explored error-correction schemes for suppressing physical noise and enabling reliable logical computation~\cite{gottesman2009introductionquantumerrorcorrection}.
Representative examples include surface codes~\cite{dennis2002topological,towards,Litinski2019gameofsurfacecodes}, color codes~\cite{colorcode}, and more general quantum LDPC codes~\cite{qLDPC}.
Building on these codes, various fault-tolerant protocols implement logical operations over encoded qubits, including code-surgery techniques for logical measurements and multi-qubit operations~\cite{Litinski2018latticesurgery,Litinski2019gameofsurfacecodes,1g44-jp62}, as well as magic-state preparation, injection, distillation, and cultivation protocols for realizing costly non-Clifford gates~\cite{Litinski2019magicstate,gidney2024magicstatecultivationgrowing}.

\paragraph{Quantum programming languages.}
Quantum programming languages span abstraction levels from circuit-construction frameworks to languages with dedicated programming and verification support.
Industrial systems such as Qiskit~\cite{qiskit}, Cirq~\cite{Cirq}, OpenQASM~3~\cite{OpenQASM3}, and Q\#~\cite{microsoft_resource_estimator,assessing} provide practical abstractions for circuits, classical control, compilation targets, and resource estimation.
Academic work has explored complementary abstractions, including formal semantics~\cite{Selinger}, uncomputation~\cite{Silq}, quantum data structures~\cite{Tower}, block encoding~\cite{Cobble}, scalable circuit generation~\cite{Quipper,ScaffCC}, and verification techniques based on predicate transformers, Hoare logics, automata, and theorem proving~\cite{dhondt2006quantum,ying2011floyd,AutoQ,QWIRE}.

\paragraph{Quantum resource estimation and fault-tolerant quantum programming.}
Quantum resource estimation is often conducted on a case-by-case basis for specific algorithms or applications.
For example, \citet{scherer2017concrete} analyze a quantum linear-systems algorithm for electromagnetic scattering, while cryptographic studies estimate the resources required for elliptic-curve discrete logarithms, Shor's algorithm, RSA factoring, and related attacks~\cite{roetteler2017quantumresourceestimatescomputing,Gidney2021howtofactorbit,babbush2026securingellipticcurvecryptocurrencies,cain2026shorsalgorithmpossible10000}.
General-purpose tools make such estimation reusable.
Qualtran~\cite{harrigan2024expressinganalyzingquantumalgorithms} provides reusable algorithmic blocks and architecture-independent counts, while Bartiq~\cite{psiquantum2024bartiq} supports symbolic and compositional estimation.
Azure QRE~\cite{microsoft_resource_estimator,vandam2023azureqre,assessing,microsoft_qdk_qrev3} maps logical programs to physical estimates under configurable hardware, error-correction, error-budget, and factory assumptions.
BenchQ~\cite{zapata_benchq} combines graph-state compilation with factory, decoder, architecture, and benchmark models; pyLIQTR~\cite{obenland_2026_18154991} reports logical-level Clifford+\(T\) estimates; and Rigetti's RRE~\cite{rigetti}, TQEC~\cite{tqec}, and the Lattice Surgery Compiler~\cite{lattice_surgery_compiler} target architecture- or surface-code-specific compilation and estimation.

Most of these tools either report only logical-level estimates or translate logical resources into physical costs under relatively fixed error-correction assumptions, with limited modeling of classical decoding overheads.
In contrast, our framework exposes programmer-visible error-correction schemes and jointly estimates space cost, time cost, and error-rate evolution from the source program, giving programmers finer control over resource usage and estimation assumptions.


\bibliography{ref}
\bibliographystyle{ACM-Reference-Format}


\end{document}